\documentclass[acmtog,nonacm]{acmart}
\authorsaddresses{}

\usepackage{booktabs} 
\usepackage{xspace}
\usepackage{array}
\usepackage[ruled]{algorithm2e} 

\SetAlFnt{\small}
\SetAlCapFnt{\small}
\SetAlCapNameFnt{\small}
\SetAlCapHSkip{0pt}

\usepackage{soul,color}
\definecolor{Red}{cmyk}{0,1,1,0}
\definecolor{Green}{cmyk}{1,0,1,0}
\definecolor{Cyan}{cmyk}{1,0,0,0}
\definecolor{Purple}{cmyk}{0.45,0.86,0,0}
\definecolor{Rosolic}{cmyk}{0.00,1.00,0.50,0}
\definecolor{Blue}{cmyk}{1.00,1.00,0.00,0}
\definecolor{Orange}{cmyk}{0,0.52,0.80,0}
\definecolor{Black}{cmyk}{1,0,0,1}

\setcopyright{none}
\copyrightyear{}
\acmDOI{}

\begin{document}
\title{Generative 3D Gaussians with Learned Density Control}

\author{Runjie Yan}
\authornote{Work done during internship at VAST}
\orcid{0009-0004-8060-6033}
\affiliation{%
 \institution{Institute for Interdisciplinary Information Sciences, Tsinghua University}
 \country{China}
}
\email{runjieyan021@gmail.com}
\author{Yan-Pei Cao}
\orcid{0000-0002-0416-4374}
\affiliation{%
 \institution{VAST}
 \country{China}
}
\email{caoyanpei@gmail.com}
\author{Peng Wang}
\orcid{0009-0009-8245-1860}
\affiliation{%
 \institution{VAST}
 \country{China}
}
\email{totoro97@outlook.com}
\author{Ding Liang}
\orcid{0000-0001-9774-4687}
\affiliation{%
 \institution{VAST}
 \country{China}
}
\email{liangding1990@163.com}
\author{Yuan-Chen Guo}
\authornote{Corresponding author}
\orcid{0000-0001-6164-8343}
\affiliation{%
 \institution{VAST}
 \country{China}
}
\email{imbennyguo@gmail.com}

\begin{abstract}
We present \textbf{Density-Sampled Gaussians (DeG)}, a novel 3D representation designed to bridge the gap between adaptive rendering primitives and scalable generative modeling. Unlike existing approaches that constrain 3D Gaussians to fixed voxel grids or arrays, DeG models Gaussian centers as samples from a learnable probability density function defined over an octree. This formulation provides a rigorous mathematical framework for \emph{adaptive density control}: by jointly optimizing the spatial density and Gaussian attributes under rendering supervision, our model naturally concentrates primitives in regions of high geometric complexity. We achieve this via a new \emph{render loss contribution gradient} that serves as a fully differentiable analogue to the discrete densification and pruning heuristics used in standard Gaussian Splatting. The resulting representation is highly flexible, supporting \emph{variable-resolution decoding} from a single latent code by simply adjusting the sampling budget. 
To enable generative synthesis, we train a latent diffusion model on DeG. We identify a critical challenge in applying diffusion to unordered set-structured latents, which can significantly slow convergence, and propose \textit{VecSeq}, a canonical re-indexing mechanism that anchors latent tokens to a deterministic 3D Sobol sequence. This transforms the ambiguous set-generation problem into a robust sequence modeling task. 
Extensive experiments demonstrate that our pipeline achieves state-of-the-art quality in single-image-to-3D generation, combining the structural adaptivity of unstructured primitives with the training stability of grid-based methods.

\end{abstract}

\begin{teaserfigure}
  \centering
  \includegraphics[width=1.0\textwidth]{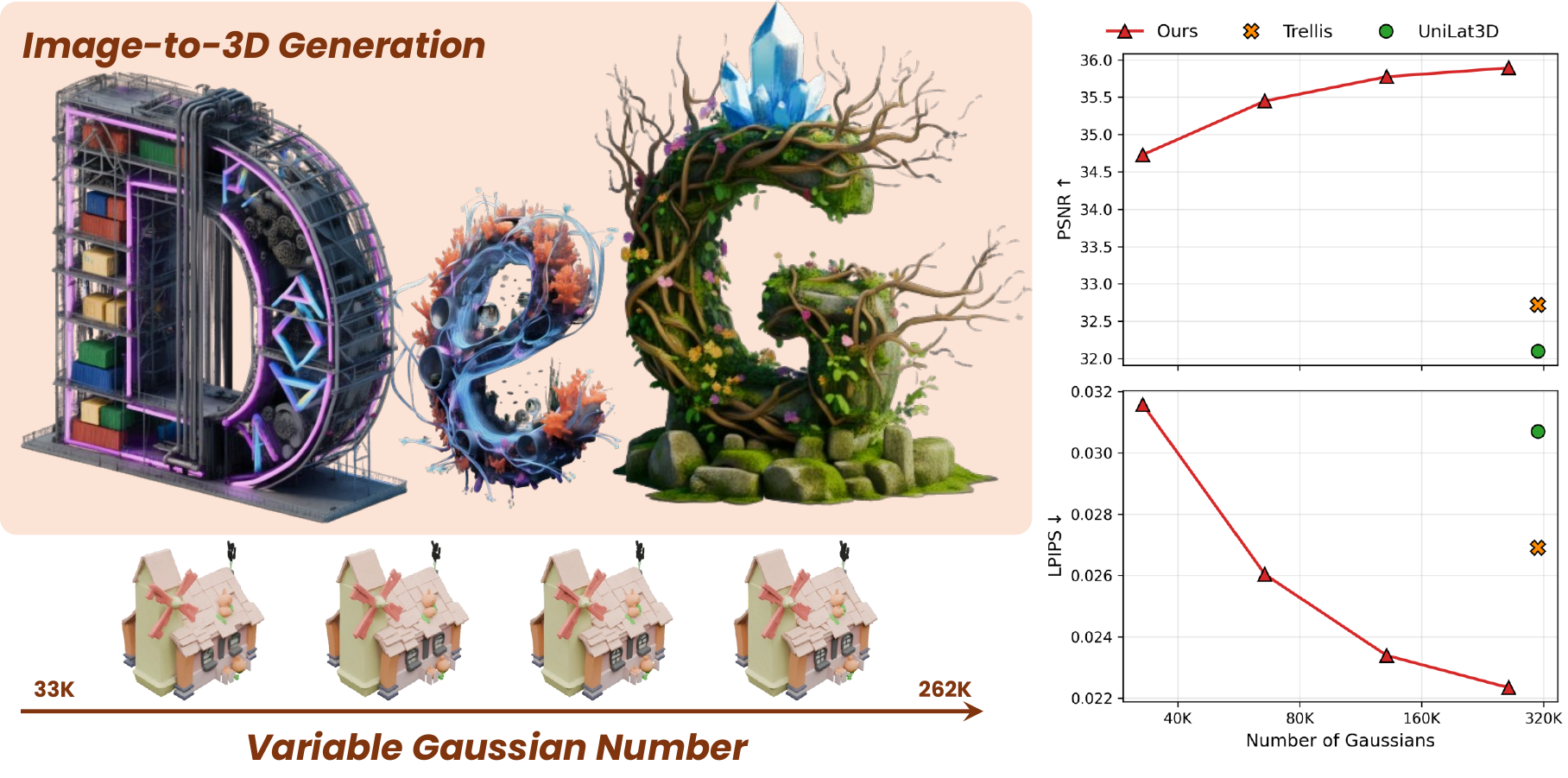}
  \caption{Teaser. Best generation samples with variable Gaussian counts, demonstrating that our model can decode an arbitrary number of 3D Gaussian splats from a single latent representation.}
  \label{fig:teaser}
\end{teaserfigure}

\ccsdesc[500]{Computing methodologies~Artificial intelligence}

%
%

\keywords{Generative Models, Gaussian Splatting}

\maketitle

\thispagestyle{empty}
\pagestyle{plain}

\newcommand{\encoder}{\mathcal{E}_\theta}
\newcommand{\probdecoder}{q_\theta}
\newcommand{\decoder}{\mathcal{D}_\theta}
\newcommand{\dit}{v_\theta}
\newcommand{\latent}{\mathcal{Z}}
\newcommand{\pointset}{\mathcal{P}}
\newcommand{\asset}{\mathcal{O}}
\newcommand{\anchorpoints}{\mathcal{P}_\text{anchor}}
\newcommand{\Lentropy}{\mathcal{L}_\text{CE}}
\newcommand{\Lrender}{\mathcal{L}_\text{render}}
\newcommand{\Lrenderbp}{\hat{\mathcal{L}}_\text{render}}
\newcommand{\Lvae}{\mathcal{L}_\text{VAE}}
\newcommand{\Lreg}{\mathcal{L}_\text{reg}}
\newcommand{\Lkl}{\mathcal{L}_\text{kl}}
\newcommand{\Lfm}{\mathcal{L}_\text{FM}}
\newcommand{\Loffset}{\mathcal{L}_\text{offset}}
\newcommand{\gs}{\mathcal{G}}

\newcommand{\probdecodertext}{stochastic density decoder\xspace}
\newcommand{\vecseq}{VecSeq\xspace}
\newcommand{\Lvecseqfm}{\mathcal{L}_\text{VecSeqFM}}
\newcommand{\ourgsvae}{DeG-VAE\xspace}
\newcommand{\lonec}{render loss contribution gradient}

\section{Introduction}
3D generative models are increasingly central to graphics and vision, enabling content creation for AR/VR, simulation, robotics, and interactive applications. A core challenge is finding a 3D representation that is both amenable to learning and capable of high-fidelity rendering at a practical cost. Recent work has explored multiple representations for generative modeling~\cite{xiang2025structured,tang2024dreamgaussian,poole2022dreamfusion}, seeking a favorable balance among expressiveness, efficiency, and differentiability. In this landscape, 3D Gaussian Splatting~\cite{kerbl20233d} has emerged as a compelling representation due to its flexibility, high visual quality, and promising rendering speed.

The quality of 3D Gaussians largely stems from the density control strategy~\cite{ye2024absgs,rota2024revising,hanson2025pup}. Iterative densification and pruning are performed throughout the fitting process to increase Gaussian density in under-fit regions and to remove Gaussians that contribute little. Density control allocates more Gaussians to complex regions and fewer to simple ones, striking a balance between the number of Gaussians and visual quality. However, densification and pruning are non-differentiable and difficult to vectorize, which makes them impractical in a generalizable learning setting. As a result, existing approaches to generative modeling of Gaussians typically represent a 3D scene with a fixed number of Gaussians tied to predefined structures. For example, GaussianCube~\cite{zhang2024gaussiancube} optimizes a fixed $N^3$ Gaussians per object and reorganizes them onto a grid using optimal transport. Structured latents~\cite{xiang2025structured,wu2025unilat3d} assign a fixed number of Gaussians to each voxel in a given sparse structure. Pixel-aligned Gaussians~\cite{zhang2024gs,xu2024grm,tang2024lgm} use a fixed number of Gaussians per image pixel or patch. None of these methods can adaptively allocate Gaussians based on local complexity, so they often require an excessive number of Gaussians to achieve high visual fidelity. This, in turn, complicates training and increases rendering cost.

In this work, we propose a generative framework that restores the adaptive capability of 3DGS without per-scene optimization. We introduce \textit{Density-Sampled Gaussians (DeG)}, a representation where Gaussian centers are dynamically sampled from a learned 3D probability density function (PDF). Rather than regressing fixed coordinates, our decoder predicts a spatial distribution indicating the likelihood of surface geometry. This formulation decouples the spatial distribution of primitives from their attributes. At inference time, we can sample an arbitrary number of anchors from this density, allowing a single trained model to generate lightweight assets for mobile applications or ultra-dense assets for high-fidelity rendering simply by varying the sample count.

The primary technical challenge lies in optimizing this stochastic density end-to-end. Since the sampling operation is non-differentiable, standard backpropagation cannot update the density based on rendering error. We address this by deriving the \textit{\lonec}, which measures the marginal contribution of each sampled anchor to the rendering loss via the difference reward~\cite{wolpert2001optimal}, and use this signal to reinforce the probability density in regions where primitives significantly reduce reconstruction error. This provides a fully differentiable alternative to the heuristic densification and pruning used in per-scene optimization.

Building on this representation, we address the generative modeling task using the latent diffusion paradigm. We encode 3D assets into a set of latent tokens and model their distribution. However, we identify a critical challenge in applying diffusion to unordered set-structured latents~\cite{zhang2024clay,li2025triposg,zhang20233dshape2vecset}: the permutation ambiguity leads to conflicting gradient signals that can significantly slow convergence and degrade generation quality. To resolve this, we propose \textit{\vecseq}, a canonical re-indexing strategy. We map the unordered latent tokens to a deterministic, low-discrepancy 3D Sobol sequence using optimal transport~\cite{berger2009optimal}. This imposes a stable spatial ordering on the latents, transforming the difficult set-generation problem into a robust sequence-generation task.

Our contributions are summarized as follows:
\begin{itemize}
\item{We introduce Density-Sampled Gaussians (DeG), a 3D representation designed for generative modeling that supports variable-sized outputs and adaptive allocation of Gaussians by sampling centers from a learnable density function.}
\item{We derive the \textit{\lonec}, an efficient signal that enables end-to-end optimization of the stochastic density function using only image reconstruction loss, effectively learning optimal primitive placement.}
\item{We propose \textit{\vecseq}, a latent re-indexing mechanism that stabilizes diffusion training on point sets by anchoring tokens to a deterministic spatial structure, achieving faster convergence and state-of-the-art generation quality.}
\end{itemize}

\begin{figure*}[t]
  \centering
  \includegraphics[width=\textwidth]{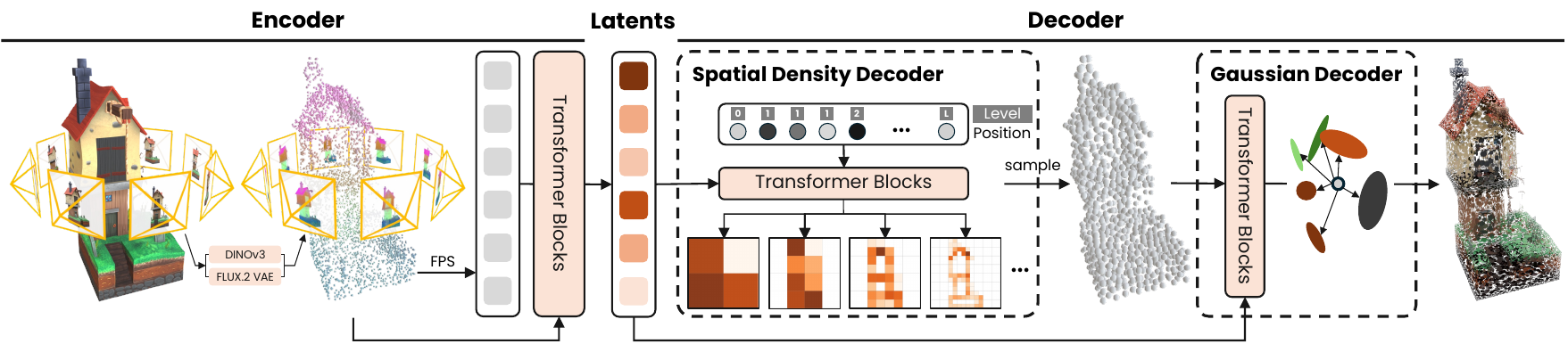}
  \caption{Overview of the DeG-VAE. Multi-view renderings are encoded using off-the-shelf feature extractors, and the features are projected to randomly sampled surface points. Points with features are encoded into latent tokens following 3DShape2VecSet~\cite{zhang20233dshape2vecset}. Taking the latent tokens as the condition, a spatial density decoder models the spatial distribution of Gaussians, and a Gaussian decoder predicts Gaussian attributes for differentiable rendering. Note that the rendering loss can be back-propagated to the spatial density decoder, allowing for adaptive density control.}
  \label{fig:pipeline}
\end{figure*}

\section{Related Work}
\label{sec:related}
\subsection{3D Gaussian Splatting}
\label{sec:background:gs}
3D Gaussian Splatting (3DGS)~\cite{kerbl20233d, yu2024mip} represents a scene or object as a set of anisotropic Gaussians.
Rendering is performed via differentiable rasterization, commonly referred to as splatting, which achieves real-time speed and high visual quality.
As a result, 3DGS has recently become popular for photorealistic rendering.
Recent improvements to 3DGS include enhanced densification~\cite{ye2024absgs,rota2024revising}, efficient training~\cite{kheradmand2024mcmc,mallick2024taming,lan20253dgs2}, sparse-view reconstruction~\cite{xiong2023sparsegs,li2024dngaussian}, and instant feed-forward inference~\cite{zhang2024gs,xu2024grm,ziwen2025long,tang2024lgm}. F4Splat~\cite{kim2026f4splat} extends densification to feed-forward reconstruction by learning heuristic densification scores.
These works aim for reconstruction and generally lack the ability to generate 3D Gaussians.

\subsection{3D Generative Models}
\label{sec:background:3dgen}
Early 3D generation methods focused on explicit 3D representations (e.g., voxels, point clouds, and meshes) and used adversarial training to model category-level shape distributions~\cite{wu2016learning,gao2022get3d}.
With the advent of large-scale 2D diffusion priors, score distillation sampling (SDS) enables 3D generation by optimizing through gradients distilled from a frozen diffusion model, without requiring curated 3D training data~\cite{poole2022dreamfusion}.
Subsequent works improve the efficiency and realism of SDS-based optimization by refining gradient formulations~\cite{yan2025consistent,wang2023prolificdreamer}, introducing additional priors~\cite{long2024wonder3d,chen2023fantasia3d,qiu2024richdreamer}, and accelerating convergence~\cite{liang2024luciddreamer,tang2024dreamgaussian}.
Despite these advances, optimization-based pipelines remain computationally expensive, motivating feed-forward large reconstruction models~\cite{hong2023lrm} that amortize reconstruction by training scalable transformers on large-scale multi-view data, enabling 3D asset prediction from images~\cite{wu2024unique3d, liu2023zero, xu2024grm, liu2023one, shi2023zero123++}.

More recently, the success of latent diffusion models (LDMs)~\cite{rombach2022high} has inspired 3D-native latent representations for scalable 3D generation.
3DShape2VecSet~\cite{zhang20233dshape2vecset} proposes a paradigm that encodes signed distance fields into an unordered set of latent tokens and performs diffusion in this latent space. More generally, unordered set generation has been explored for 3D point sets~\cite{fan2017point}.
CLAY~\cite{zhang2024clay} further develops this direction with scaled-up data processing and training, achieving large-scale asset generation.
Notable follow-ups along this line include TripoSG~\cite{li2025triposg}, Hunyuan 2.1~\cite{hunyuan3d2025hunyuan3d}, and Direct3D~\cite{wu2024direct3d}, among others. While LATTICE~\cite{lai2025lattice} further extends the set generation paradigm with a two-stage coarse-to-fine pipeline.
Another branch of latent 3D generation seeks to improve geometric detail via sparse voxel hierarchies~\cite{ren2024xcube} or sparse structures~\cite{li2025sparc3d,xiang2025structured,wu2025unilat3d}.
These works primarily target surface geometry generation, and relatively few focus on generating 3D Gaussians.

\subsection{Generation of 3D Gaussians}
We focus on generating high-quality 3D Gaussians.
Simply combining Gaussian representations with optimization-based pipelines (e.g., DreamGaussian~\cite{tang2024dreamgaussian}) is often insufficient, as performance is bounded by the 2D vision priors and the optimization remains costly.
Recent works explore feed-forward generation to improve efficiency, while existing approaches are typically constrained by their neural output parameterization.
Structured-latent methods~\cite{wu2025unilat3d,xiang2025structured} assign a fixed number of Gaussians to each voxel in a sparse 3D structure.
Pixel-aligned lifting approaches~\cite{zhang2024gs,xu2024grm,tang2024lgm}, inspired by LRM- or VGGT-style pipelines~\cite{wang2025vggt}, predict a fixed number of Gaussians per pixel or patch.
Due to these architectural constraints, such methods struggle to preserve the key advantage of 3D Gaussians: a highly flexible representation that can adaptively allocate capacity to important regions.
GaussianCube~\cite{zhang2024gaussiancube} attempts to recover output flexibility by constructing an optimal-transport mapping between structured grids and a target set of Gaussians, where the targets are obtained by per-object Gaussian fitting.
This introduces substantial training overhead: generating ground-truth Gaussians via per-object fitting is time-consuming, and the method still typically produces a fixed number of Gaussians due to the underlying grid resolution.
AtlasGaussian~\cite{yang2024atlas} represents 3D Gaussians by sampling from learned UV patches.
While it can generate arbitrarily many primitives in principle, it samples uniformly within each patch, which does not model a global adaptive distribution and limits representational flexibility.
MaskGaussian~\cite{liu2025maskgaussian} treats Gaussians as probabilistic entities via predicted masks, but still operates on a fixed spatial scaffold without learning a global density.

\section{Method}
\label{sec:method}

\subsection{Overview}
\label{sec:method:overview}

\begin{figure*}[t]
  \centering
  \resizebox{0.98\linewidth}{!}{%
    \begin{tabular}{@{}c@{\hspace{0.05cm}}c@{\hspace{0.05cm}}c@{\hspace{0.05cm}}c@{}}
      \begin{tabular}{@{}c@{}}
        \includegraphics[height=5cm]{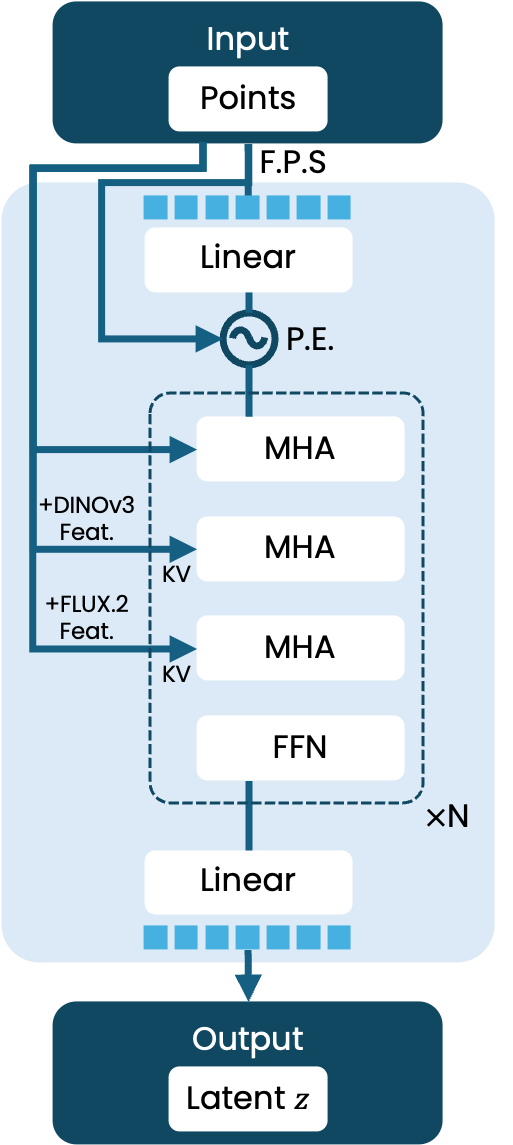}\\[-0.8ex]
        \scriptsize (a) Encoder $\encoder$
      \end{tabular}
      &
      \begin{tabular}{@{}c@{}}
        \includegraphics[height=5cm]{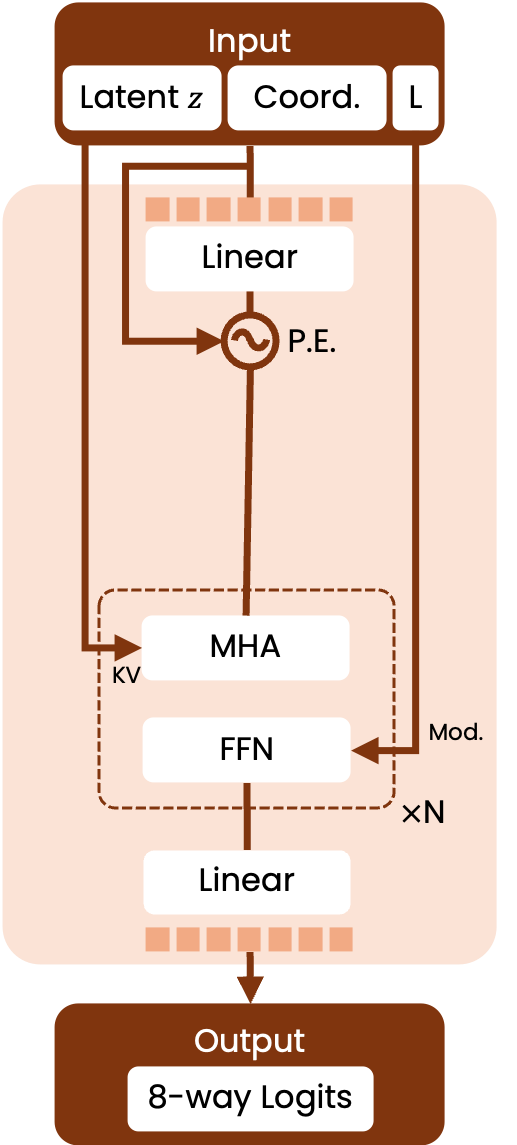}\\[-0.8ex]
        \scriptsize (b) Density Decoder $\probdecoder$
      \end{tabular}
      &
      \begin{tabular}{@{}c@{}}
        \includegraphics[height=5cm]{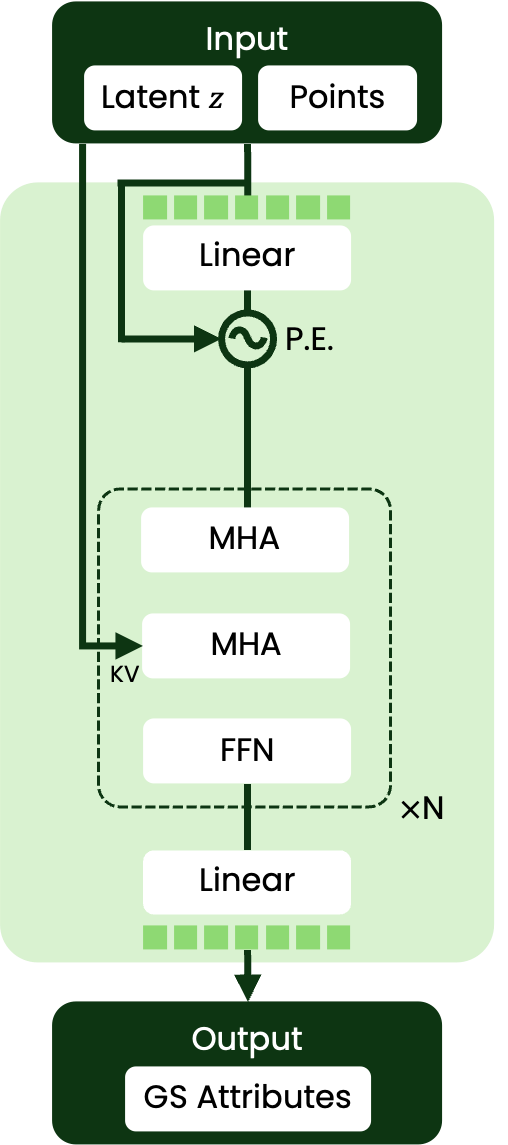}\\[-0.8ex]
        \scriptsize (c) GS Decoder $\decoder$
      \end{tabular}
      &
      \begin{tabular}{@{}c@{}}
        \includegraphics[height=5cm]{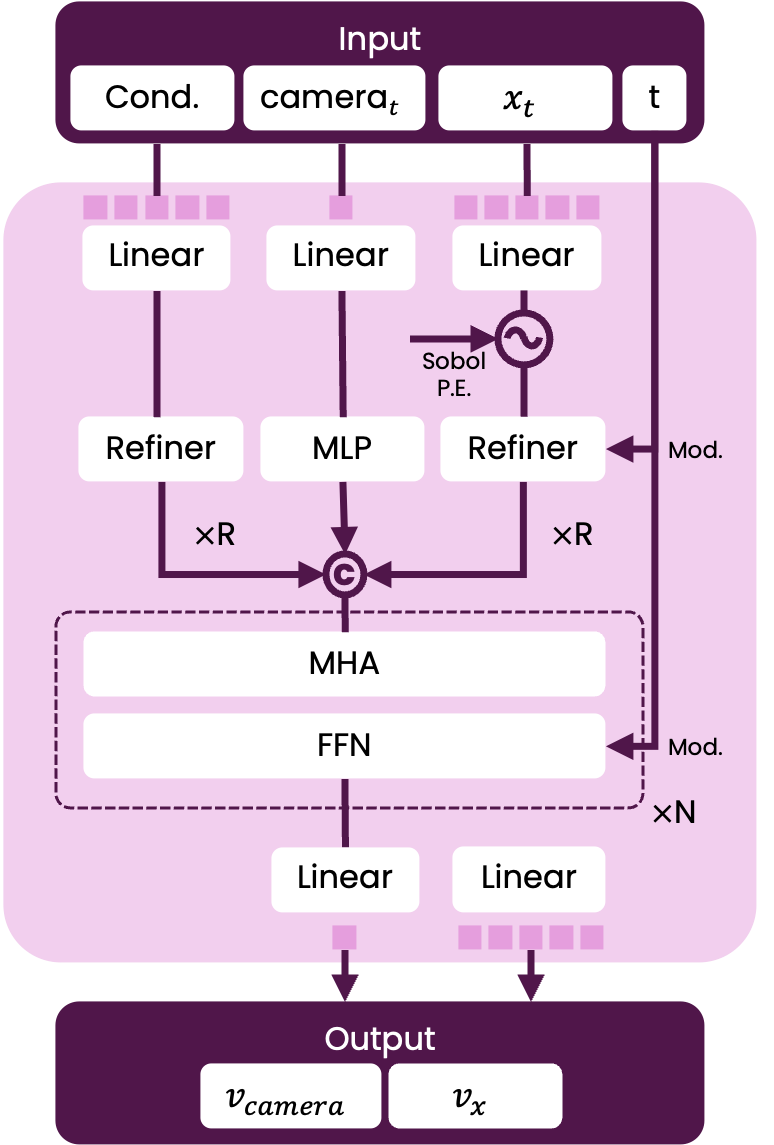}\\[-0.8ex]
        \scriptsize (d) VecSeq DiT $\dit$
      \end{tabular}
    \end{tabular}%
  }
  \caption{Detailed architecture for encoding, decoding, and generation. The refiners in (d) are transformer layers used for processing inputs from different modalities following S3-DiT~\cite{cai2025z}. To enable RoPE~\cite{su2024roformer} for multi-modal input in the DiT, we use a lightweight head to predict the 3D index given the latent in each DiT layer, following RePo~\cite{li2025repo}.}
  \label{fig:architecture}
\end{figure*}

Our pipeline aims to bridge the gap between fixed-structure generative models and the adaptive nature of 3DGS.
While we follow the latent diffusion paradigm~\citep{rombach2022high}, we diverge from approaches that constrain Gaussians to regular fixed-size structures~\citep{zhang2024gaussiancube, xiang2025structured}.
Instead, we introduce a representation that naturally supports variable resolution and adaptive allocation.
Our method consists of two core components:
(1) A \textit{Density-sampled Gaussian VAE (DeG-VAE)}, which encodes 3D assets into a compact latent space and decodes them via a learned spatial probability density. This allows the model to allocate Gaussian primitives dynamically and to train the VAE end-to-end with a novel density-aware multi-view rendering loss (Sec.~\ref{sec:method:gsv} and Sec.~\ref{sec:method:render-aware}). 
(2) A \textit{VecSeq} diffusion transformer, which models the distribution of these latent tokens and is trained to recover the latents conditioned on a single input image. To address the convergence challenge of diffusion on unordered sets, VecSeq introduces a canonical re-indexing mechanism based on optimal transport, enabling robust and scalable generation (Sec.~\ref{sec:method:diff}).
Figure~\ref{fig:pipeline} provides a high-level overview of the pipeline, while Figure~\ref{fig:architecture} illustrates the detailed neural architectures of these components. We detail each component in the following subsections.

\begin{figure*}[t]
  \centering
  \includegraphics[width=\textwidth]{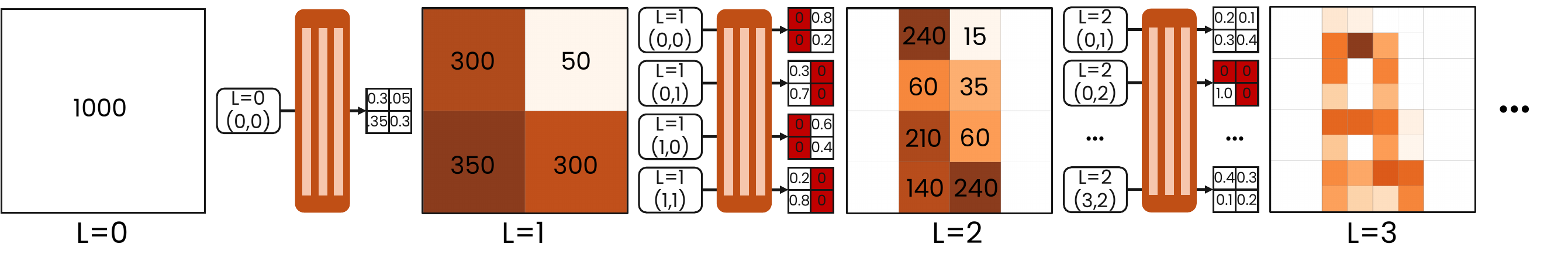}
  \caption{A 2D illustration of the point sampling process. Starting from the coarsest level, the network iteratively predicts the density value for each occupied voxel at the current level until the finest level is reached. Integers in the grid denote the number of points allocated to each voxel, given a target of 1,000 points for point sampling.}
  \label{fig:sample}
\end{figure*}

\subsection{Density-sampled Gaussian VAE}
\label{sec:method:gsv}
The core of our representation is the decoupling of \emph{geometric distribution} (i.e., where primitives exist) from \emph{primitive attributes} (i.e., appearance and local shape).

\paragraph{Set Encoder}
For a 3D asset $\asset$, we represent its geometry and appearance as a set of latent tokens $\latent=\{z_i \in \mathbb{R}^C \}_{i=1}^{M}$, adhering to the set-latent paradigm~\cite{zhang20233dshape2vecset}.
To capture high-fidelity details, we aggregate information from both multi-view RGB renderings and explicit surface geometry.
We render $K$ views of $\asset$ given camera poses $\{\pi_k\}_{k=1}^{K}$ and extract feature maps using DINOv3~\cite{simeoni2025dinov3} for semantic consistency and a FLUX.2 VAE~\cite{flux-2-2025} for high-frequency texture details.
Simultaneously, we sample a dense point cloud $\pointset=\{p_i\}_{i=1}^{N}$ from the asset surface. Following TRELLIS~\cite{xiang2025structured}, we project each point $p_i$ onto the multi-view feature maps and average the retrieved features across all views (occlusions are not handled, following prior work~\cite{xiang2025structured}). This yields two complementary feature-augmented point sets:
\begin{equation}
\label{eqn:gsv-project-dinov3}
\pointset^{\mathrm{dinov3}}=\{f_i \in \mathbb{R}^{C_1}, p_i \in \mathbb{R}^3 \}_{i=1}^{N_1},
\end{equation}
\begin{equation}
\label{eqn:gsv-project-flux2}
\pointset^{\mathrm{flux2}}=\{f_i \in \mathbb{R}^{C_2}, p_i \in \mathbb{R}^3 \}_{i=1}^{N_2},
\end{equation}
We compress these variable-length point features into a fixed-size latent set $\latent$ using a transformer-based set encoder $\encoder$:
\begin{equation}
\label{eqn:gsv-set-encode}
\latent=\encoder (\mathrm{FPS}(\pointset)|\pointset^{\mathrm{dinov3}},\pointset^{\mathrm{flux2}}),
\end{equation}
where $\mathrm{FPS}$ denotes Farthest Point Sampling, selecting $M$ representative centers to seed the encoder attention.

\paragraph{Stochastic Density Decoding}
Standard 3D decoders typically map latent tokens to a fixed number of primitives or a uniform voxel grid. This ignores a fundamental property of 3DGS: visual quality depends on the \emph{adaptive} concentration of primitives in regions of high geometric or textural complexity~\cite{kerbl20233d, ren2025fastgs, mallick2024taming}.
To bake this adaptivity into the generative model, we formulate Gaussian center prediction as a sampling process from a learned conditional probability density $\probdecoder(x \mid \latent)$ over $\mathbb{R}^3$.
At inference time, we draw $P$ \emph{anchor points} from this density:
\begin{equation}
\anchorpoints = \{x_i\}_{i=1}^{P}\sim \probdecoder(\cdot\mid \latent).
\end{equation}
Crucially, $P$ is not fixed by the architecture; it can be adjusted at inference time to trade off rendering speed for fidelity.

\paragraph{Efficient Octree-Based Sampling}
Defining $\probdecoder$ over a dense voxel grid is computationally prohibitive ($O(N^3)$). Instead, we parameterize the density using an $L$-level octree factorization, enabling an effective resolution of $(2^L)^3$ while maintaining sparse computation. Let $x_{0:l}$ denote the index of an octree cell at level $l$ along the path to $x$. We factorize the joint probability as:
\begin{equation}
\label{eqn:octree-factor}
\probdecoder(x\mid \latent)=\prod_{l=1}^{L} \probdecoder(x_{0:l}\mid x_{0:l-1},\latent),
\end{equation}
where each term represents an 8-way categorical distribution over the children of a parent cell.
Each conditional $\probdecoder(x_{0:l}\mid x_{0:l-1},\latent)$ is implemented as a shared transformer $\theta$ that cross-attends to the latent tokens $\latent$ and outputs 8 logits for the active parent cell.
We implement this via efficient ancestral sampling (details in Supplementary). We maintain a frontier of \emph{active cells} containing samples. At each level, we only evaluate the probability logits for active cells, routing samples to children based on $\probdecoder$. Empty branches are naturally pruned, and the process repeats until level $L$. This yields discrete leaf indices, which are dequantized into continuous anchor positions $\anchorpoints$ via uniform sampling within the leaf volume. A 2D illustration of the sampling process is shown in Fig.~\ref{fig:sample}.

\paragraph{Attribute Decoding and Local Expansion}
With the sampled anchors $\anchorpoints$ establishing the spatial support of the representation, the renderable geometry and appearance are resolved in a subsequent stage.
Given the anchors and global latents $\latent$, we employ a transformer-based attribute decoder to predict the parameters of the Gaussian primitives (opacity, scaling, rotation, and spherical harmonic coefficients). To further capture local surface details, we implement a local expansion mechanism: each anchor $x_i$ spawns $K$ individual Gaussians with learned local offsets:
\begin{equation}
\label{eqn:gsv-attr-decode}
\{\{g_i^k\}_{k=1}^{K}\}_{i=1}^{P}=\decoder(\anchorpoints \mid \latent),
\end{equation}
where $\decoder$ is a learnable transformer-based attribute decoder, and the Gaussian position $x_i^k$ is predicted by adding an offset to the anchor $x_i$. 
This hierarchical approach, i.e., global density sampling followed by local expansion, allows the model to represent large uniform areas with few anchors while densely populating complex details, yielding $N = P \cdot K$ total splats.
\subsection{Differentiable Density Optimization}
\label{sec:method:render-aware}

A key challenge in our pipeline is optimizing the spatial density $\probdecoder$. Standard methods rely solely on structural supervision (e.g., cross-entropy against surface voxels), which often misaligns with rendering needs: allocating too many primitives to flat, textured surfaces and too few to thin geometric structures.
Ideally, we want to update the density based on \emph{rendering} feedback. However, because anchor locations $\anchorpoints$ are samples from $\probdecoder$, the rendering loss $\Lrender$ is not directly differentiable with respect to the density parameters $\theta$.
To bridge this gap, we derive the \emph{\lonec} that backpropagates rendering feedback into the probabilistic density, effectively performing ``differentiable densification and pruning.''

\paragraph{Structural Initialization}
We first anchor the density using explicit geometry. Given the target distribution $p(x)$ derived from surface points, we minimize the cross-entropy loss over the octree structure:
\begin{align}
\label{eqn:struct-ce}
\Lentropy
&= -\sum_{x_{0:L}} p(x_{0:L}) \log \probdecoder(x_{0:L}\mid \latent) \\
&= -\sum_{x_{0:L}} p(x_{0:L}) \sum_{l=1}^{L} \log \probdecoder(x_{0:l}\mid x_{0:l-1},\latent) \\
&= -\sum_{l=1}^{L} \sum_{x_{0:l}} p(x_{0:l}) \log \probdecoder(x_{0:l}\mid x_{0:l-1},\latent),
\end{align}
where $x_{0:L}$ denotes a level-$L$ leaf cell, $p(x_{0:L})$ is the normalized histogram of surface points assigned to leaves, and $p(x_{0:l})$ denotes its marginal distribution over level-$l$ cells.

\paragraph{Rendering Supervision}
For appearance supervision, we sample camera poses $\pi$ and minimize image reconstruction losses between $\mathcal{R}(\mathcal{G},\pi)$ and the target images, where $\mathcal{R}$ is the differentiable Gaussian splatting rendering function and $\mathcal{G}$ is the set of decoded 3D Gaussian primitives.
We minimize the weighted sum of L1 loss, SSIM loss and LPIPS loss:
\begin{equation}
\label{eqn:render-loss}
\Lrender
= \mathcal{L}_\text{l1} + \lambda_\text{ssim} \mathcal{L}_\text{ssim} + \lambda_\text{lpips} \mathcal{L}_\text{lpips}.
\end{equation}

\paragraph{Backpropagating Rendering to Density}
Unlike prior works~\cite{xiang2025structured} that treat structure and appearance as separate optimization problems, we unify them. We seek to minimize the expected rendering loss over the density distribution.
Specifically, we also propagate rendering supervision to the structural density \probdecodertext $\probdecoder$ and regard the structural loss $\Lentropy$ mainly as a regularizer.
The loss gradient with respect to anchors $\anchorpoints$, which are sampled from the decoded densities, cannot be directly propagated to VAE parameters $\theta$.
Fortunately, we note that the gradient of the expectation of the rendering loss with respect to the density distribution can be computed:
\begin{align}
\nabla_{\theta} \Lrender 
&= \nabla_{\theta} \mathbb{E}_{x_i \sim \probdecoder} \left[ \Lrender (\anchorpoints = \{x_i\}_{i=1}^{P}) \right]\\
&= \mathbb{E}\left[
  \mathcal{L}(\{x_i\}_{i=1}^{P}) \nabla_{\theta} \log\left(\prod_{j=1}^{P} \probdecoder(x_j)\right)
\right]\\
&= \mathbb{E}\left[
  \sum_{j=1}^{P} \mathcal{L}(\{x_i\}_{i=1}^{P}) \nabla_{\theta} \log(\probdecoder(x_j))
\right] \label{eqn:policy-gradient}\\
&= \mathbb{E}\left[
  \sum_{j=1}^{P} (\mathcal{L}(\{x_i\}_{i=1}^{P}) - \mathcal{L}(\{x_i\}_{i \neq j}^P)) \nabla_{\theta} \log(\probdecoder(x_j))
\right], \label{eqn:difference-reward}
\end{align}
where Eq.~\ref{eqn:policy-gradient} corresponds to the standard policy-gradient~\cite{sutton1999policy}, and Eq.~\ref{eqn:difference-reward} can be interpreted as an advantage estimation, also known as difference reward~\cite{wolpert2001optimal,tumer2007distributed}.
Here, $\{x_i\}_{i \neq j}^P$ denotes the anchor set excluding $x_j$. The difference term $\Delta \Lrender = \Lrender(\{x_i\}_{i=1}^{P}) - \Lrender(\{x_i\}_{i \neq j}^P)$ measures how much anchor $x_j$ decreases the rendering loss;
Intuitively, this term increases the probability density at locations where the presence of anchor $x_j$ leads to a larger reduction in rendering error.

\paragraph{Efficient \lonec}

Directly evaluating Eq.~\ref{eqn:difference-reward} remains impractical, since each leave-one-out baseline would require an additional rendering pass for every sampled anchor.
Our key observation is that, for the pixel-wise $\mathcal{L}_{\text{l1}}$ term in $\Lrender$, the standard 3DGS backward rasterization already maintains the transmittance and accumulated back color needed to estimate the loss change caused by removing a primitive.
We therefore accumulate primitive-level contributions inside the same CUDA backward pass at negligible overhead, sum them over primitives from the same anchor, and directly backpropagate the resulting anchor-level signal to $\log\!\left(\probdecoder(x_j\mid \latent)\right)$.
This fused computation relies on the additive per-pixel structure of $\mathcal{L}_{\text{l1}}$; full primitive-level derivations and CUDA implementation details are provided in the supplementary material.
Finally, our VAE model is trained with a combination of structural supervision and rendering supervision via:
\begin{equation}
\label{eqn:vae-loss}
\Lvae = \lambda_\text{struct} \Lentropy + \lambda_\text{render} (\Lrender + \Lrenderbp) + \lambda_\text{reg} \Lreg + \lambda_\text{kl} \Lkl,
\end{equation}
where $\Lrenderbp$ denotes the additional \lonec described above, and $\Lreg$ is a regularization term on predicted GS parameters (details are provided in the Supplementary). 

We optimize this objective using a three-stage curriculum.
\begin{description}
\item[\textit{Stage 1 (Structural Initialization):}] We optimize $\encoder$ and $\probdecoder$ using only $\Lentropy$ to establish a coarse geometric hull, analogous to standard 3DGS initialization (e.g., SfM-derived). This prevents degenerate solutions (e.g., zero-opacity collapse) from random Gaussians, and takes only $\sim$6\% of total training time.
\item[\textit{Stage 2 (Appearance):}] We train the attribute decoder $\decoder$ with a small gaussian count and large batch size using $\Lrender + \Lentropy$, locking in appearance. This accelerates convergence and is optional.
\item[\textit{Stage 3 (Joint Refinement):}] We train all parameters end-to-end with the full $\Lvae$, enabling $\Lrenderbp$ to provide signals for density reallocation. We also randomize the number of anchors $P$ to encourage the model to generalize across different resolution budgets.
\end{description}

\begin{figure}[t]
  \centering
  \includegraphics[width=\linewidth]{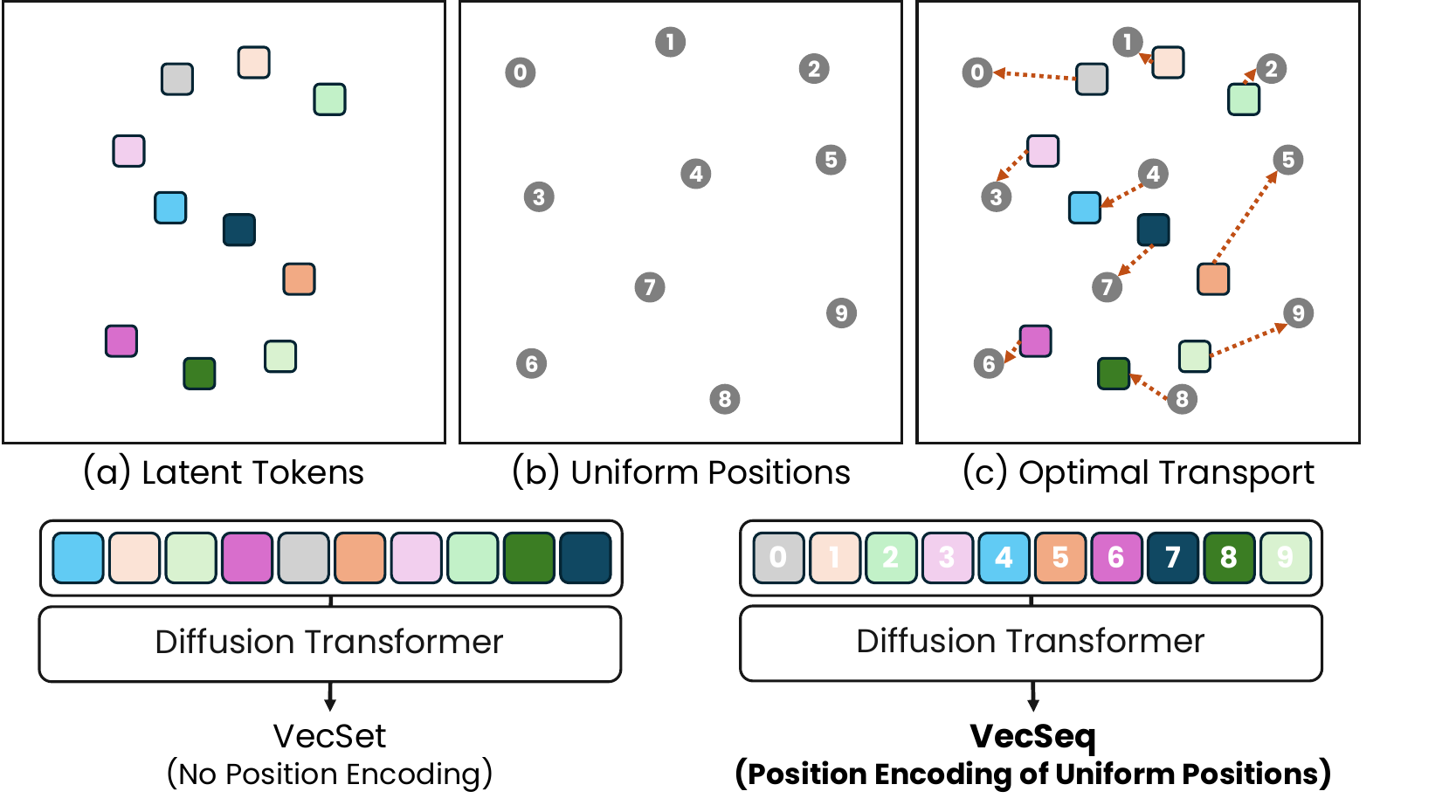}
  \caption{\vecseq re-indexing. Latent tokens are associated with 3D positions during encoding and then canonically ordered by matching them to deterministic 3D Sobol anchors, turning an unordered latent set into a stable vector sequence for diffusion denoising.}
  \label{fig:vecseq}
  \vspace{20 pt}
\end{figure}

\subsection{\vecseq Diffusion}
\label{sec:method:diff}
We model the distribution of latent codes $\latent$ using a diffusion transformer.
We adopt the Flow Matching framework~\cite{lipman2022flow} with an S3-DiT backbone~\cite{cai2025z}.
The training objective is
\begin{equation}
\label{eqn:flow-matching}
\Lfm = \mathbb{E}_{t,x_0,\epsilon} || \dit (x_t,t) - (\epsilon - x_0) ||^2_2,
\end{equation}
where $x_t$ is the latent state at time $t$, interpolated between data $x_0$ and noise $\epsilon$. To recover the conditioning-view pose, we jointly predict a compact noisy camera token $c_t$ and concatenate it to $x_t$ during training. Details of $c_t$ are provided in the supplementary material.

\paragraph{The Permutation Ambiguity}
A fundamental challenge in training diffusion models on latent \textit{sets} is permutation invariance. Our set encoder produces an unordered set of tokens $\latent=\{z_i\}_{i=1}^{M}$. 
Unlike pixels in an image, which have fixed coordinates, set tokens have no intrinsic ordering. If we feed these unordered sets directly to a diffusion model, the pairing between noise tokens and data tokens becomes arbitrary ($M!$ possible pairings). The model is forced to learn an average over all permutations, resulting in slow convergence and blurry, mode-averaged generations.

\paragraph{\textbf{\vecseq}: Canonical Serialization via Optimal Transport}
To resolve this, we propose \emph{\vecseq}, a method to transform the unordered latent set into a canonically ordered vector sequence.
While the latent tokens $z_i$ themselves lack coordinates, they are derived from surface points $p_i \in \mathrm{FPS}(\pointset)$ during encoding (Eq.~\ref{eqn:gsv-set-encode}). We could use these $p_i$ to sort the tokens, but these positions are asset-specific and unknown at inference time.

Instead, we align the tokens to a \emph{fixed, deterministic} spatial structure that is shared across all assets. We choose a 3D Sobol sequence~\cite{sobol1967distribution} $\mathcal{S} = \{s_j\}_{j=1}^{M}$ as our anchor structure. Sobol sequences are low-discrepancy quasi-random sequences that cover the unit cube $[0,1]^3$ more uniformly than standard random sampling, ensuring a balanced spatial scaffold.
During training, we compute an optimal assignment $\pi^\star$ that matches the asset-specific FPS points $\{p_i\}$ to the fixed Sobol anchors $\{s_j\}$ by minimizing the total transport cost:
\begin{equation}
\label{eqn:vecseq-ot}
\pi^{\star}=\mathrm{3D\;OT\;Assign}(\{p_i\},\{s_j\}).
\end{equation}
We then reorder the latent tokens according to this map, yielding a sequence $\tilde{\latent} = \{\tilde{z}_j\}_{j=1}^{M}$ where $\tilde{z}_j = z_{\pi^{\star}(j)}$.
This assignment is computed once as an offline preprocessing step, incurring zero cost during training or inference.

Crucially, this reordering associates the $j$-th token of \emph{any} asset with the spatial region around the $j$-th Sobol anchor $s_j$.
We inject this spatial prior into the diffusion model by adding a sinusoidal positional embedding of $s_j$ to the $j$-th token.
At inference time, the model simply predicts a sequence of length $M$, knowing implicitly that the $j$-th output corresponds to the spatial location $s_j$.
This effectively converts the difficult set-generation problem into a stable sequence-generation problem, significantly improving convergence and fidelity.

Conceptually, this assignment is similar to GaussianCube~\cite{zhang2024gaussiancube}, which also uses OT to assign Gaussians to a fixed cube structure; however, our assignment is performed over latent tokens rather than directly over Gaussian primitives. Moreover, using a single universal template, i.e., the Sobol anchors $\{s_j\}$, allows each object to be matched independently, yielding \emph{linear} $O(N)$ complexity, unlike classical permutation synchronization methods~\cite{huang2013consistent} that require pairwise matching with $O(N^2)$ complexity and are intractable for large open-vocabulary datasets.

\begin{figure}[t]
  \centering
  \includegraphics[width=\linewidth]{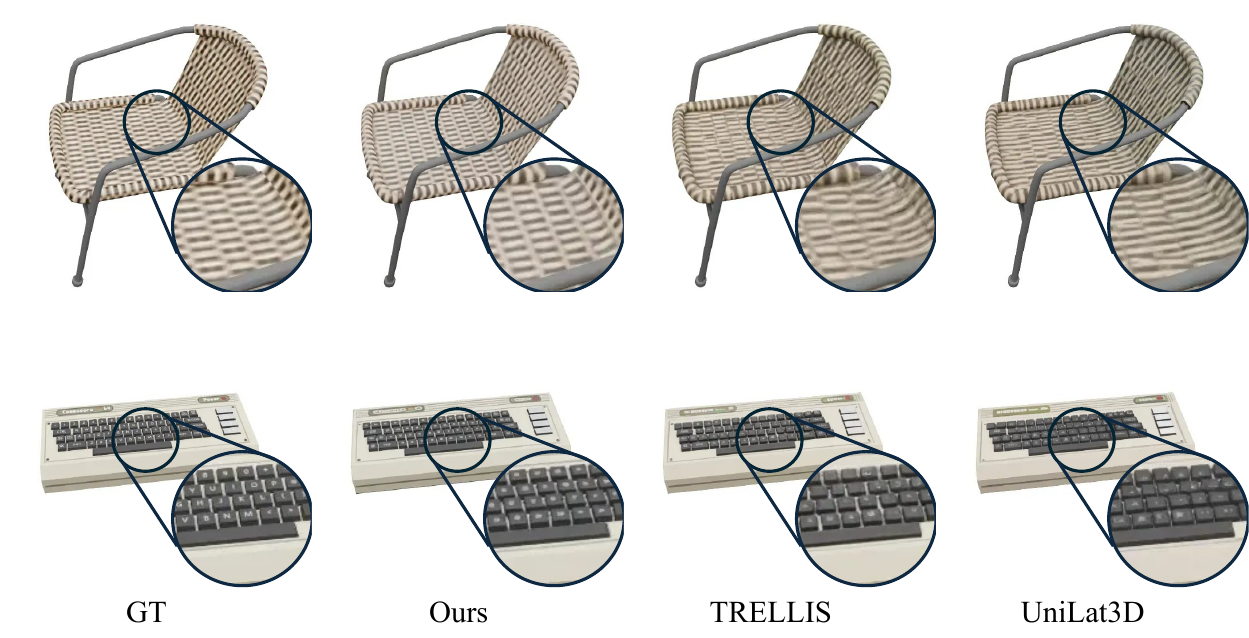}
  \caption{Qualitative comparison of 3D reconstruction. Under a matched Gaussian budget (\ourgsvae with 262K vs. baselines with $\approx 310$K), our model achieves higher visual fidelity. As shown in the zoom-in views, \ourgsvae preserves fine details and complex structures significantly better than the baselines.}
  \label{fig:vae-compare}
  \vspace{15 pt}
\end{figure}
\section{Experiments}
\label{sec:experiments}

We evaluate both the reconstruction component (\ourgsvae) and the generative component (latent diffusion on \vecseq). We report quantitative metrics and qualitative comparisons.

\subsection{Implementation Details}
\label{sec:experiments:impl}
We train \ourgsvae and the latent generation model on the Objaverse~\cite{deitke2022objaverse} and Objaverse-XL~\cite{deitke2023objaverse} subsets of the TRELLIS-500K dataset~\cite{xiang2025structured}.
In the first stage of VAE training,  we use $1024$ latent tokens and use cross entropy loss to supervise $8192$ points.
In the second stage of VAE training, we randomly sample $1024$ -- $8192$ tokens and $2048$ anchors for GS rendering, corresponding to $N=2048 \cdot K$ final Gaussians after local expansion.
In the third stage of VAE training, we randomly sample $1024$ -- $8192$ tokens and $1024$ -- $8192$ anchors per asset for rendering (the two quantities are not necessarily identical).
We train the VAE for approximately 10 days on 32 NVIDIA A800 GPUs and the flow-matching model for approximately 11 days on 32 NVIDIA A800 GPUs.

For quantitative evaluation, we use the Toys4K dataset~\cite{stojanov2021using} for both reconstruction and generation; this test set is unseen during our model training. For qualitative results, we use a set of high-quality, self-collected images for image-conditioned generation.

\subsection{Reconstruction Results}
\label{sec:experiments:recon}

\begin{figure}[t]
  \centering
  \begin{minipage}[t]{0.51\linewidth}
    \centering
    \includegraphics[width=\linewidth]{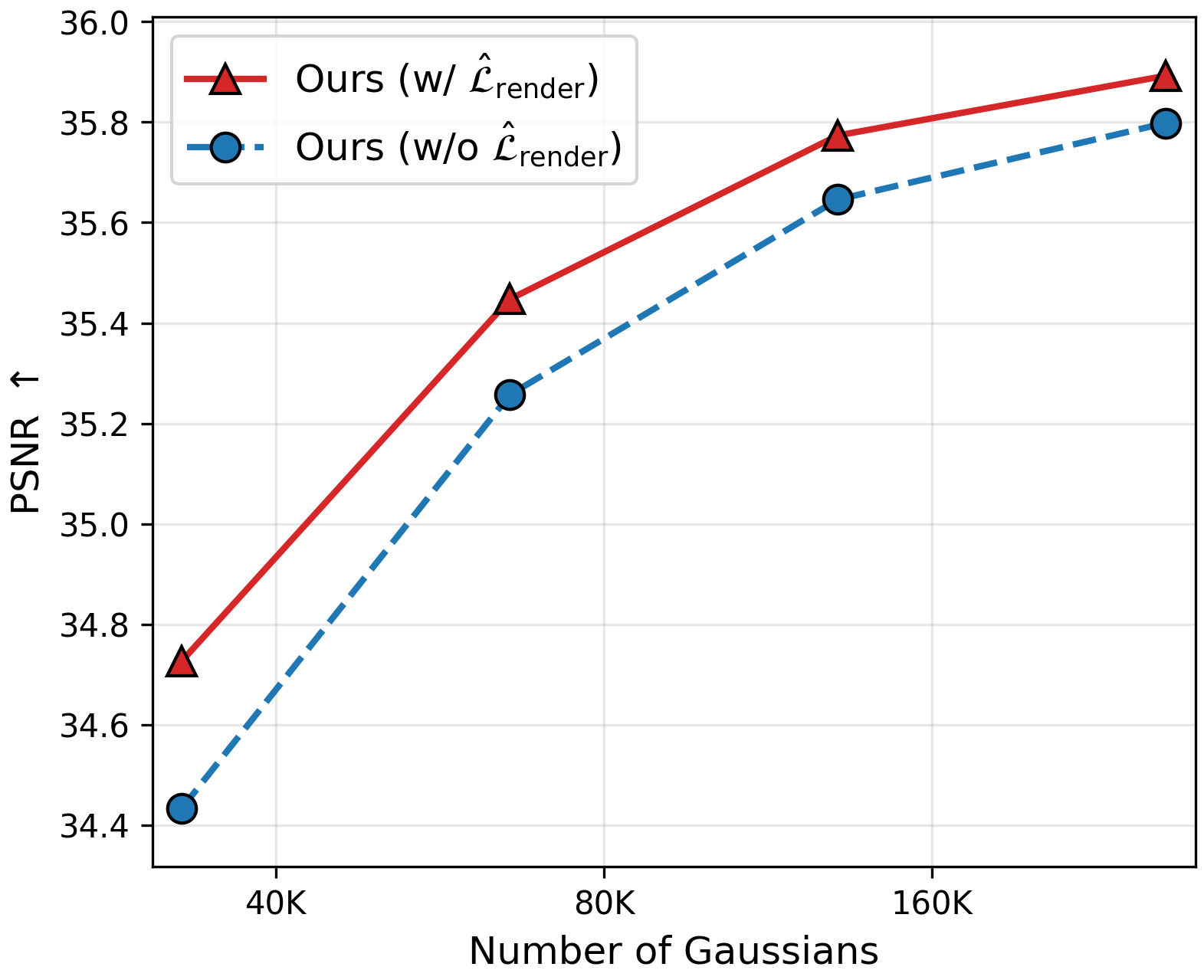}

    \small (a)
  \end{minipage}
  \hfill
  \begin{minipage}[t]{0.4786\linewidth}
    \centering
    \includegraphics[width=\linewidth]{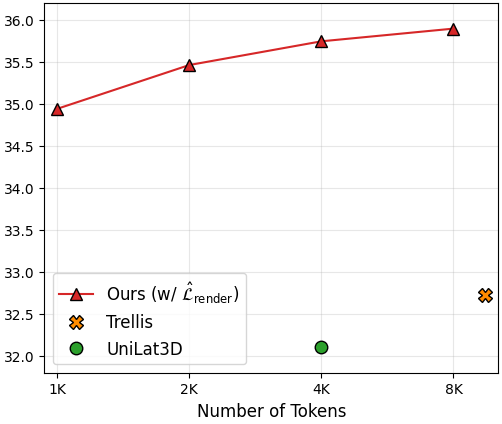}

    \small (b)
  \end{minipage}
  \caption{Quantitative evaluation. PSNR versus (a) the number of decoded Gaussians and (b) the latent token length, comparing TRELLIS, UniLat3D, our method, and our method without the \lonec ($\Lrenderbp$).}
  \vspace{20 pt}
  \label{fig:exp-scaling}
\end{figure}

\begin{figure*}[t]
  \centering
  \includegraphics[width=\textwidth]{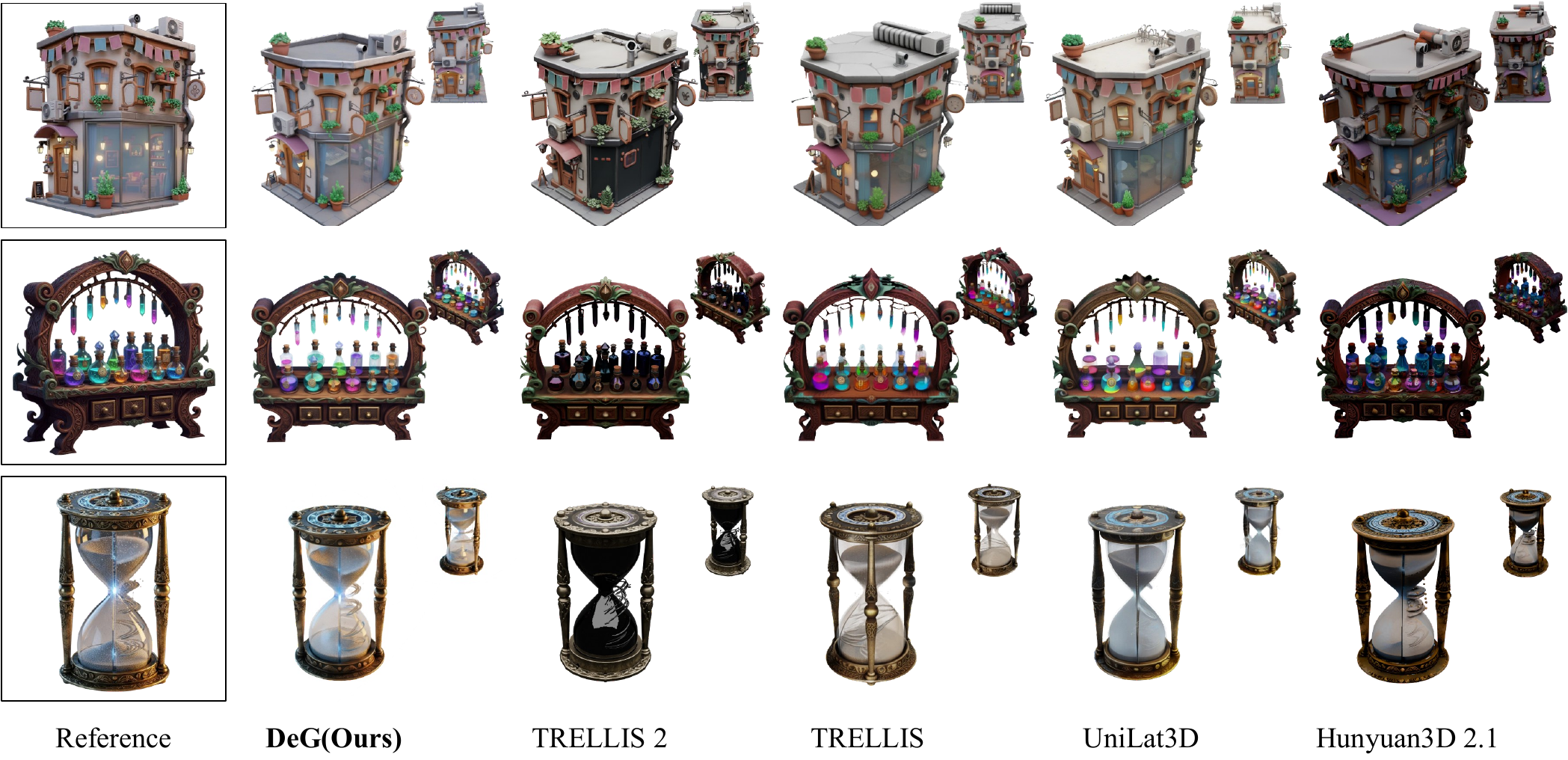}
  \caption{Generation comparison. We compare our generated 3D Gaussian assets with representative textured 3D generation models under the same rendering settings. Our model can generate 3D Gaussians with accurate structure and appearance details.}
  \label{fig:gen-comparison}
  \vspace{15 pt}
\end{figure*}
\subsection{Generation Results}
\label{sec:experiments:bp}

\begin{table*}[t]
  \centering
  \caption{Quantitative reconstruction results on the Toys4K dataset. We compare our \ourgsvae against baselines in terms of PSNR$\uparrow$, SSIM$\uparrow$, and LPIPS$\downarrow$. Our method consistently outperforms competing approaches under a comparable Gaussian budget. Dec. records the decoding time per object on a single NVIDIA 4090 GPU (batch size=4).}
  \label{tab:gsv-recon}
  \begin{tabular}{l|ccc|ccc}
    \toprule
    Method &$ \# \text{Token}_{(\# \text{Dim})}$ & \# Gaussians & Dec. (s) & PSNR$\uparrow$ & SSIM$\uparrow$ & LPIPS$\downarrow$ \\
    \midrule
    TRELLIS   & $\approx 9679_{(\approx 77k)}$ & $\approx 310k$
    & 0.034 & 32.72 & 0.9734 & 0.0269 \\
    UniLat3D  & $4096_{(32k)}$ & $\approx 310k$
    & 0.074 & 32.10 & 0.9715 & 0.0307 \\
    Ours      & $8192_{(131k)}$ & $262k$
    & 0.10 & \textbf{35.89} & \textbf{0.9787} & \textbf{0.0223} \\
    \bottomrule
  \end{tabular}
  \vspace{15 pt}
\end{table*}

We evaluate the reconstruction performance of \ourgsvae on Toys4K. For each object, we render 16 viewpoints and compute image-level metrics between the ground-truth renderings and renderings from the decoded Gaussians. We report PSNR, SSIM, and LPIPS, and compare against representative baselines including TRELLIS~\cite{xiang2025structured} and UniLat3D~\cite{wu2025unilat3d}. As shown in Table~\ref{tab:gsv-recon} and Fig.~\ref{fig:vae-compare}, under a comparable Gaussian budget, \ourgsvae substantially outperforms all competitors across PSNR, SSIM, and LPIPS. TRELLIS and UniLat3D assign a fixed number of Gaussians to each voxel, which inevitably overspends capacity in simple regions while under-allocating it in complex ones. In contrast, \ourgsvae learns to allocate Gaussian density directly from rendering supervision, using the available Gaussians more effectively to maximize visual fidelity.

\paragraph{Variable-Sized Gaussians.}
A key advantage of the DeG representation is its ability to produce variable-sized Gaussian sets. This flexibility enables explicit trade-offs between rendering/memory cost and visual quality by varying the number of sampled anchors $P$, which determines the final Gaussian count $N=P K$ after local expansion. In Fig.~\ref{fig:exp-scaling} (a), we analyze reconstruction quality as a function of $N$. Performance improves steadily as the Gaussian count increases. Notably, \ourgsvae reaches the same visual quality (LPIPS) as TRELLIS while using less than 1/2 as many Gaussians.

\paragraph{Learned Density Control.}
To validate the effectiveness of optimizing density via rendering supervision with $\Lrenderbp$, we train a Stage-3 VAE variant that disables $\Lrenderbp$ while keeping all other settings and training steps fixed. Fig.~\ref{fig:exp-scaling} reports reconstruction quality with and without $\Lrenderbp$ across different decoded Gaussian counts. Incorporating $\Lrenderbp$ consistently improves reconstruction, with the largest gains appearing in the low-budget regime, consistent with the intuition that adaptive allocation is most valuable when capacity is limited. We provide qualitative visualizations of this effect, including comparisons of the generated anchor point clouds, in the supplementary material.

\paragraph{Token Length.}
Latent token length controls the degree of compression in our representation. In Fig.~\ref{fig:exp-scaling} (b), we visualize reconstruction quality as a function of token length and observe consistent improvements as more tokens are used, highlighting the favorable scaling behavior of DeG.

\begin{table*}[t]
  \centering
  \caption{3DGS generation metrics. CLIP-I measures image-level cosine similarity between rendered and prompt images. We report FD$\downarrow$, KD$\downarrow$, and CLIP-I$\uparrow$ on rendered multi-view images. The best results are shown in bold, and the second-best results are underlined.}
  \label{tab:gen-metrics}
  \begin{tabular}{l|c|ccccc}
    \toprule
    Method & Repr. & CLIP-I$\uparrow$ & $\text{FD}_{\text{incep}}\downarrow$ & $\text{KD}_{\text{incep}}\downarrow$& $\text{FD}_{\text{dinov2}}\downarrow$ & $\text{KD}_{\text{dinov2}}\downarrow$  \\
    \midrule
    Hunyuan3D 2.1     
    & Mesh & 89.09 & 4.86 & \underline{0.05} & 95.55 & 7.99\\
    TRELLIS-2         
    & Mesh & 89.83 & 4.55 & \underline{0.05} & 63.36 & 2.74 \\  
    \midrule
    LGM               
    & 3DGS & 80.49 & 35.7 & 1.74 & 626.4 & 112.4 \\
    DiffusionGS       
    & 3DGS & 84.82 & 15.7 & 0.49 & 287.9 & 34.52\\
    TRELLIS        
    & 3DGS & 91.68 & \underline{2.55} & \textbf{0.02} & \underline{33.39} & \textbf{1.3}\\ 
    UniLat3D        
    & 3DGS & \underline{91.69} & 2.72 & \textbf{0.02} & 33.71 & \textbf{1.3} \\ 
    Ours
    & 3DGS & \textbf{92.26} & \textbf{2.42} & \textbf{0.02} & \textbf{31.16} & \underline{1.4} \\
    \bottomrule
  \end{tabular}
  \vspace{15 pt}
\end{table*}

\paragraph{Quantitative comparison.}
We measure Gaussian generation performance using image-condition alignment (CLIP-I) and distributional metrics computed on rendered multi-view images ($\text{FD}_{\text{incep}}$, $\text{KD}_{\text{incep}}$, $\text{FD}_{\text{dinov2}}$, and $\text{KD}_{\text{dinov2}}$). We detail the computation of these scores in the supplementary material. We compare against representative baselines, including mesh generation models (Hunyuan3D 2.1~\cite{hunyuan3d2025hunyuan3d}, TRELLIS-2~\cite{xiang2025native}) and Gaussian generation models (GaussianAnything~\cite{lan2024gaussiananything}, LGM~\cite{tang2024lgm}, DiffusionGS~\cite{cai2025baking}, TRELLIS~\cite{xiang2025structured}, UniLat3D~\cite{wu2025unilat3d}).
Table~\ref{tab:gen-metrics} shows that our method achieves the highest image-conditioning alignment score and delivers the best performance on most distributional metrics, demonstrating state-of-the-art 3D generation with strong visual consistency.

\paragraph{Qualitative comparison.}
We present qualitative comparisons in Fig.~\ref{fig:gen-comparison}. Our method excels in both generation quality and image-prompt alignment compared with prior approaches, producing higher-fidelity 3D Gaussian results with more detailed geometry and texture. In addition, our generations better match prompt colors and preserve fine-grained details in the corresponding object parts.

\label{sec:experiments:ablation}

\begin{table}[t]
  \centering
  \caption{Reordering ablation. Both variants use the same encoder and decoder weights trained with VecSet-style unordered latents; the only difference is whether Sobol-anchor positional embeddings (PE) are added to the reordered tokens in diffusion training. The diffusion model is trained for the same number of steps (80K). $\text{KD}_{\text{dinov2}}\downarrow$ is reported $100\times$.}
  \label{tab:vecseq-ablation}
  \begin{tabular}{l|ccc}
    \toprule
    & CLIP-I $\uparrow$ & $\text{FD}_{\text{dinov2}}\downarrow$ & $\text{KD}_{\text{dinov2}}\downarrow$ \\
    \midrule
    w/o reordering (VecSet)  & 89.39 & 75.08 & 4.79 \\
    \textbf{w/ reordering (Ours)}  & \textbf{90.01} & \textbf{66.94} & \textbf{3.81} \\
    \bottomrule
  \end{tabular}
  \vspace{15 pt}
\end{table}

\paragraph{Effect of Token Reordering (\vecseq vs. VecSet).}
We compare \vecseq against a VecSet-style baseline that uses the same encoder and decoder weights and differs only in whether \vecseq reordering is applied during diffusion training. Without reordering, the Sobol-anchor positional embeddings carry no consistent spatial meaning: the same index can correspond to different geometric features across objects, effectively reducing the model to a VecSet-style baseline. In contrast, \vecseq reordering via OT makes each token index consistently correspond to the same spatial region, allowing the positional encoding to become informative. Table~\ref{tab:vecseq-ablation} shows that reordering improves both prompt alignment and distributional quality.

\section{Conclusions}

In this work, we presented Density-Sampled Gaussians (DeG), a generative 3D Gaussian representation that replaces non-differentiable densification and pruning with a learnable, rendering-optimized density defined over an octree. By sampling Gaussian centers from this density, DeG supports variable-sized outputs and adaptive allocation of Gaussians to locally complex regions, enabling favorable trade-offs between fidelity and rendering cost.
We further introduced a paired learning pipeline that trains an autoencoder to compress 3D assets into compact latent tokens and decode them into DeG, with density optimized end-to-end under rendering supervision. Our experiments demonstrate that this design translates into substantially improved reconstruction quality under a comparable Gaussian budget $N=P K$, and that the \lonec ($\Lrenderbp$) provides consistent gains, especially in low-budget regimes where smart allocation matters most. DeG also exhibits strong scaling behavior: reconstruction improves smoothly as either the sampled anchor count $P$ (and therefore the final Gaussian count $N$) or the latent token length increases.
Finally, we addressed a key convergence challenge in generative modeling with vector-set latents arising from permutation ambiguity in diffusion training. Our proposed VecSeq formulation assigns token positions via optimal transport to enable positional encoding, leading to faster convergence and higher-quality single-image conditional generation. Together, these contributions establish a practical and scalable foundation for high-fidelity 3D Gaussian generation and open new opportunities for controllable, resource-aware 3D content synthesis.

\paragraph{Limitations and Future Work.}
Despite these strengths, DeG has several limitations.
First, as a single-image-to-3D method, back-facing regions are not observed during inference and may exhibit lower quality; failure-case visualizations and analysis are provided in the supplementary material.
Second, our representation currently only targets 3DGS rather than meshes; however, because the latent space faithfully reconstructs both 3D shape and texture, it might already contain the necessary information for a textured mesh decoder, making direct textured mesh generation a promising direction for future work.

\begin{acks}

This work was supported in part by the International (Hong Kong, Macao, and Taiwan) Collaborative R\&D Project, Beijing Major Science and Technology Project under Contract No.Z251100007125016.
\end{acks}

\begin{figure*}[p]
  \centering
  \includegraphics[width=1.0\textwidth]{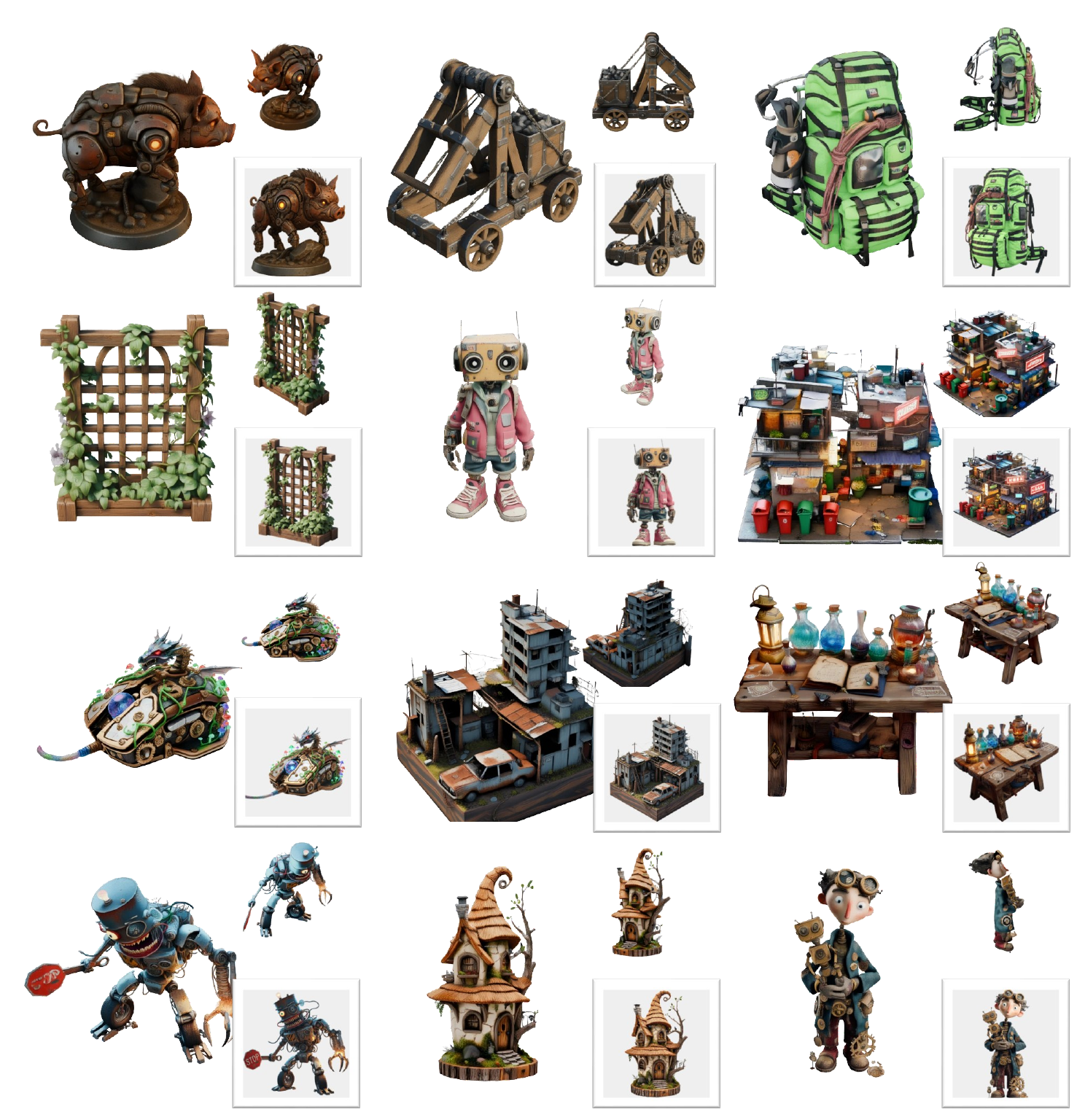}
  \caption{Generated samples. A full-page gallery of diverse 3D Gaussian generations from our model, rendered from two different viewpoints.}
  \label{fig:generated-samples}
\end{figure*}

\cleardoublepage

\bibliographystyle{ACM-Reference-Format}
\bibliography{sample-bibliography}

\clearpage                                                                                                                                                                   
\appendix                                                                                                                                                                    
\begin{center}                                                                                                                                                               
{\LARGE\bfseries Supplementary Material}\\[0.5em]                                                                                                                            
{\large Generative 3D Gaussians with Learned Density Control}                                                                                                                
\end{center}                                                                                                                                                                 
\vspace{1em}

\section{Evaluation Metrics}
\label{sec:appendix:metrics}

To quantitatively evaluate the quality of our generated 3D assets, we evaluate 2D renderings.
For each generated asset, we render images from $8$ uniformly distributed azimuth angles at a fixed elevation of $30^\circ$. The images at $45^\circ$ are used as the reference images (the left-front view).
We use a resolution of $512\times 512$ for all renderings.
We compare the distribution of these generated images against a reference set of images from the ground-truth test set.

To ensure a fair rendering comparison across representations and baselines, we normalize the rendered object scale before evaluation. For Gaussian assets, we set the camera radius, defined as the distance from the camera to the generated object center, to $2\times$ the maximum bounding-box extent. This is equivalent to scaling the Gaussian bounding box to $1$ and rendering with radius $2$. For mesh assets, including both ground-truth meshes and meshes produced by baselines, we normalize the object scale to $1$ and render with radius $2$. We then render all assets using a field of view of $40^\circ$.

\paragraph{CLIP-Score}
To measure the alignment between the generated 3D assets and the input image prompts, we use the CLIP-Score metric.
We employ the pre-trained CLIP ViT-L/14 model~\cite{radford2021learning} to extract embeddings for both the rendered images and the text prompts.
The score is calculated as the average cosine similarity between the rendered-image and reference-image embeddings across all rendered views and test samples.
A higher CLIP-Score indicates better semantic alignment.

\paragraph{Distributional Metrics (FD and KD)}
We employ Fr\'echet Distance (FD) and Kernel Distance (KD) to assess the fidelity and diversity of the generated images compared with the real reference distribution.
We use all rendered images in the Toys4K test dataset as the reference distribution, forming a reference image set of size $4,000\times 8=32,000$. The generated image set also has size $4,000\times 8=32,000$.
We compute these metrics using two different feature extractors:
\begin{itemize}
    \item \textbf{InceptionV3} ($\text{FD}_{\text{incep}}$, $\text{KD}_{\text{incep}}$): We use the standard InceptionV3 network~\cite{szegedy2016rethinking} pretrained on ImageNet. These metrics (equivalent to FID and KID) focus on the perceptual quality and high-level semantics of the images.
    \item \textbf{DINOv2} ($\text{FD}_{\text{dinov2}}$, $\text{KD}_{\text{dinov2}}$): We use the DINOv2 ViT-L/14 model~\cite{oquab2023dinov2} to extract features. DINOv2 features are known to capture more robust geometric and structural information, providing a complementary assessment to Inception-based metrics. We use the CLS token feature as the feature representation for each image.
\end{itemize}

\begin{algorithm}[!t]
\SetAlgoLined
\KwIn{Latent code $\latent$, total samples $P$, max level $L$}
\KwOut{Set of 3D anchor points $\anchorpoints$, log-probabilities $\mathcal{L}$}
\tcp{Initialize active frontier with root cell}
Initialize active frontier $\mathcal{F}_0 \leftarrow \{(\text{root}, P, 0)\}$\;
\For{$l \leftarrow 1$ \KwTo $L$}{
    $\mathcal{F}_l \leftarrow \emptyset$\;
    \ForEach{active cell $(c_{parent}, n_{parent}, \log p_{parent}) \in \mathcal{F}_{l-1}$}{
        \tcp{Predict child distribution from latent code}
        Compute logits: $h \leftarrow \text{Model}(c_{parent}, \latent)$\;
        Compute probabilities: $D \leftarrow \text{Softmax}(h)$\;
        Compute log-probabilities: $\log D \leftarrow \text{LogSoftmax}(h)$\;
        \tcp{Distribute parent samples to children}
        Distribute parent count $n_{parent}$ into children counts $\{n_0, \dots, n_7\}$ according to $D$ with \emph{systematic sampling}\;
        \For{$k \leftarrow 0$ \KwTo $7$}{
            \If{$n_k > 0$}{
                \tcp{Update cumulative log-probability}
                $\log p_{child} \leftarrow \log p_{parent} + \log D[k]$\;
                $\mathcal{F}_l \leftarrow \mathcal{F}_l \cup \{(c_{parent} \cdot 8 + k, n_k, \log p_{child})\}$\;
            }
        }
    }
}
$\anchorpoints \leftarrow \emptyset$; $\mathcal{L} \leftarrow \emptyset$\;
\tcp{Convert leaf indices to continuous coordinates}
\ForEach{leaf cell $(c_{leaf}, n_{leaf}, \log p_{leaf}) \in \mathcal{F}_L$}{
    Determine spatial bounds $B$ of cell $c_{leaf}$\;
    Sample $n_{leaf}$ points $\{x^{(j)}\}_{j=1}^{n_{leaf}} \sim \text{Uniform}(B)$\;
    $\anchorpoints \leftarrow \anchorpoints \cup \{x^{(j)}\}_{j=1}^{n_{leaf}}$\;
    $\mathcal{L} \leftarrow \mathcal{L} \cup \{\log p_{leaf}\}_{j=1}^{n_{leaf}}$\;
}
\Return $\anchorpoints, \mathcal{L}$
\caption{Efficient Batched Octree Sampling}
\label{alg:sampling}
\end{algorithm}

\section{Camera Tokens}
In DiT training, the model jointly predicts a camera token $c_t$ for each object to reconstruct the camera pose of the conditioning image. Because the training camera is randomly sampled and always points toward the center of the object, we represent the camera pose using a 5D vector. The first three components of $c_t$ encode the unit viewing direction, the fourth component is the reciprocal of the camera-to-center distance $d$, and the fifth component is the camera scale, computed as $d \cdot \tan(\mathrm{fov}/2)$. Once the object is generated, this 5D camera latent is sufficient to recover the full camera pose. However, because the conditioning image is always scaled according to the alpha bounding box during both training and inference, following prior work~\cite{xiang2025structured}, the predicted camera scale/distance may be inaccurate.

\begin{figure*}[t]
  \centering
  \setlength{\tabcolsep}{1pt}
  \renewcommand{\arraystretch}{0.95}
  \makebox[\textwidth][c]{%
  \begin{tabular}{ccccccc}

    \rotatebox[origin=l]{90}{%
      \parbox[c]{0.15\textwidth}{%
        \centering
        Condition
      }%
    } &
    \includegraphics[height=0.15\textwidth]{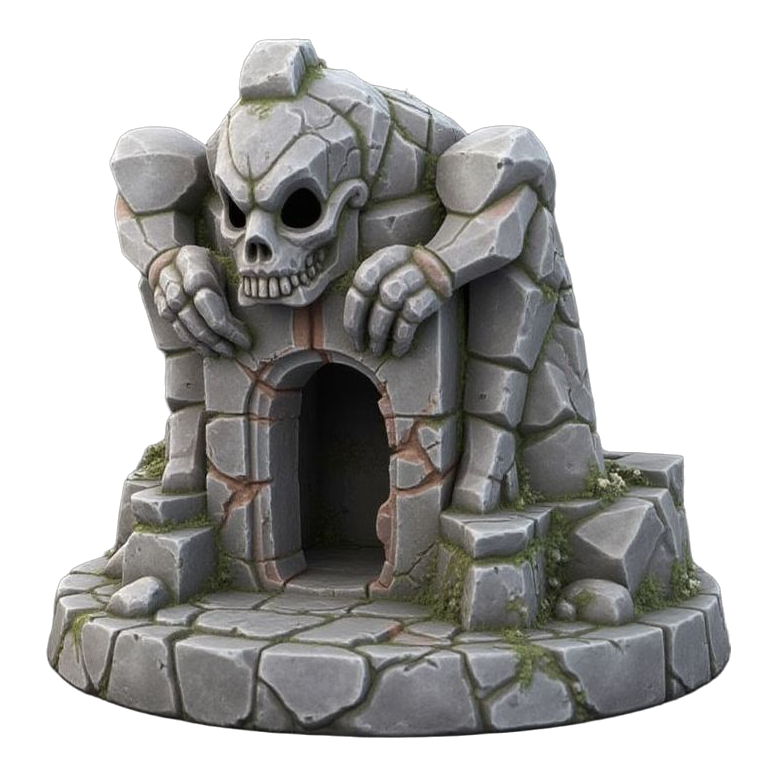} &
    \includegraphics[height=0.15\textwidth]{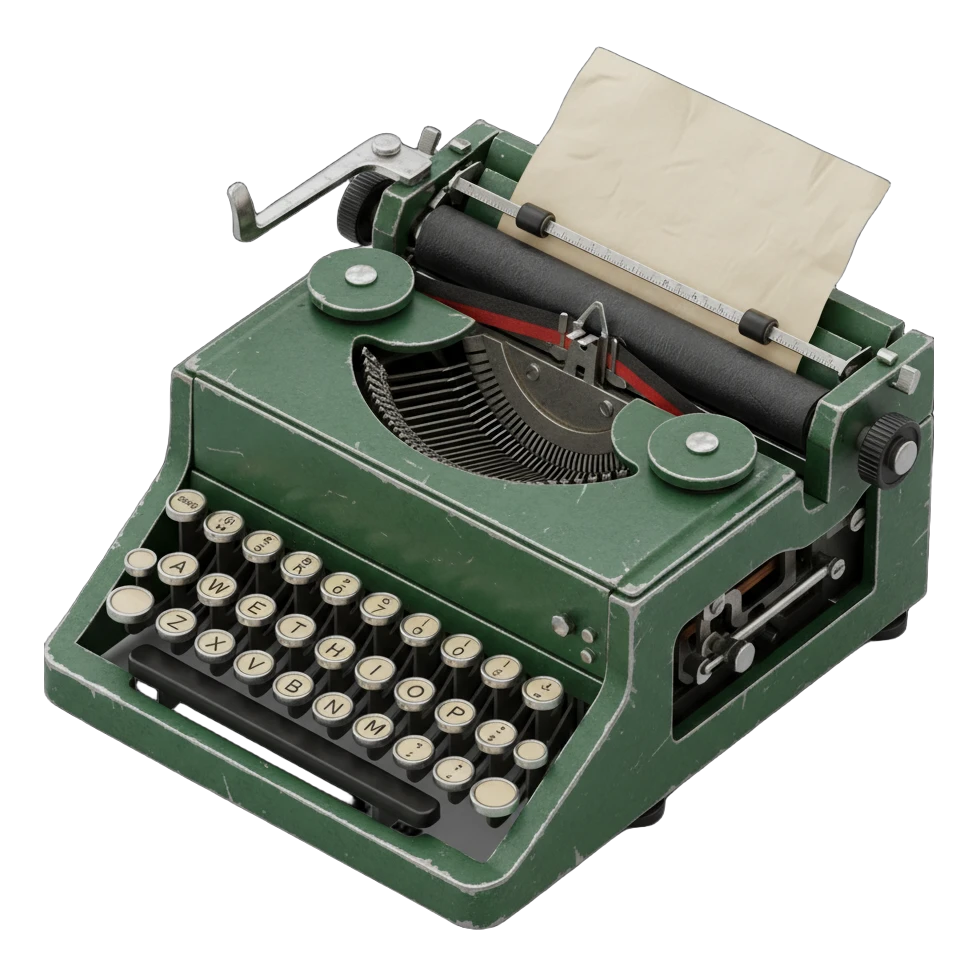} &
    \includegraphics[height=0.15\textwidth]{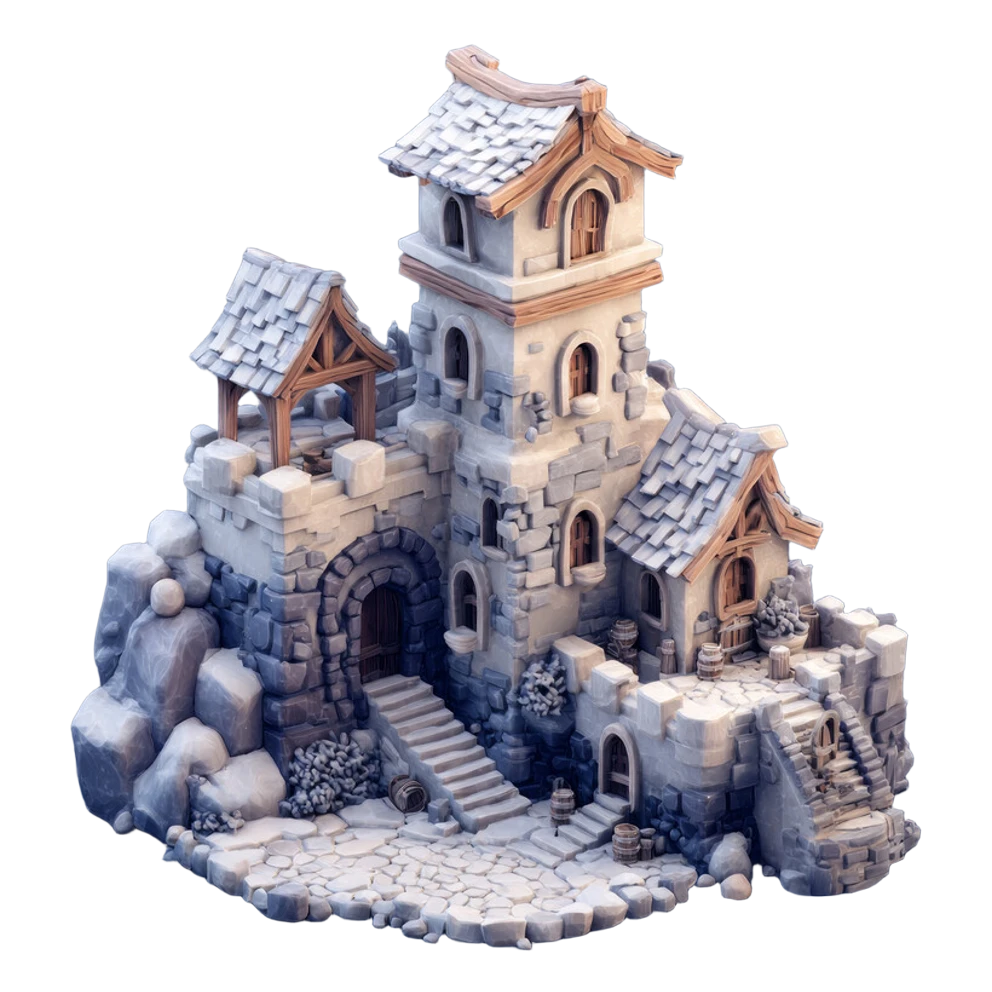} &
    \includegraphics[height=0.15\textwidth]{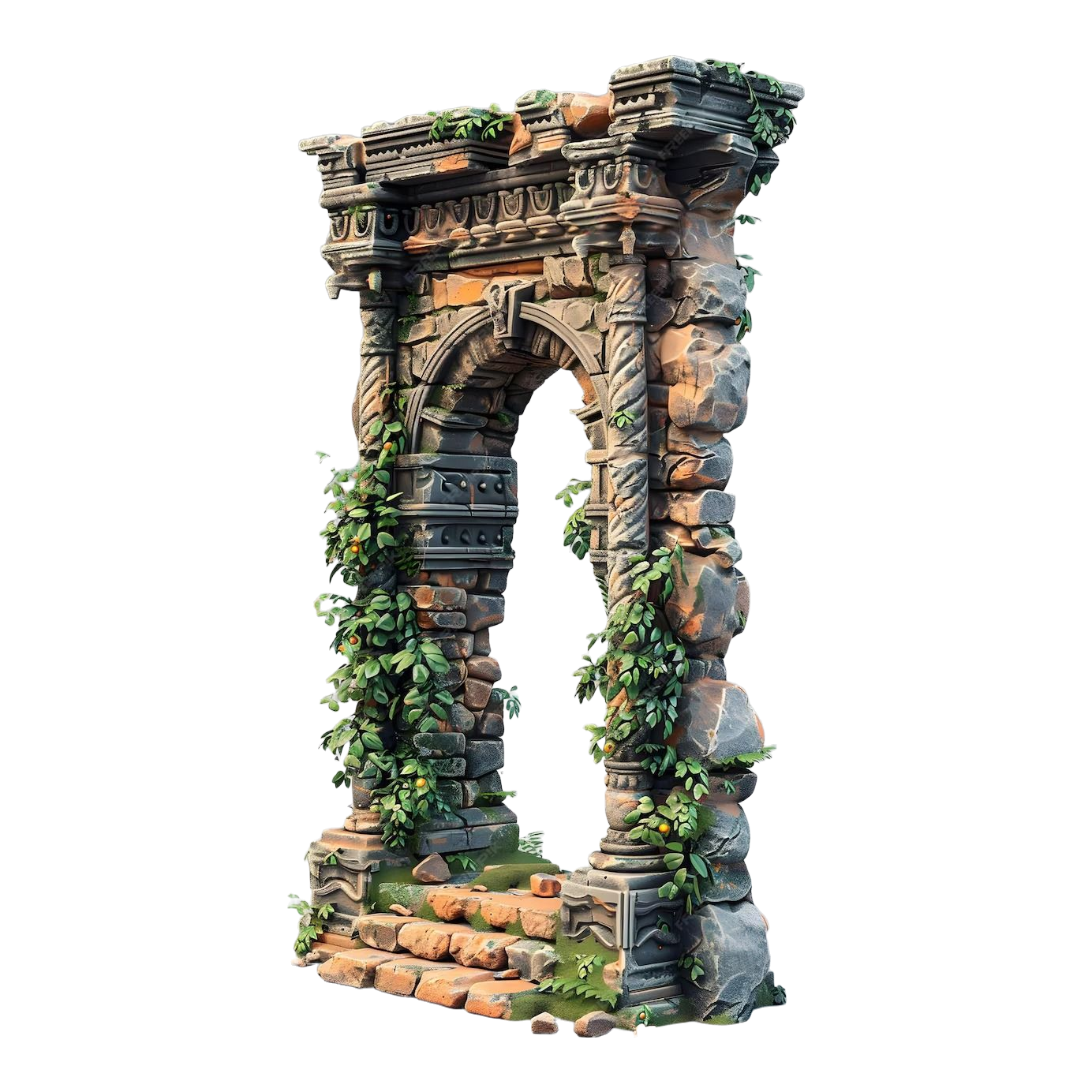} &
    \includegraphics[height=0.15\textwidth]{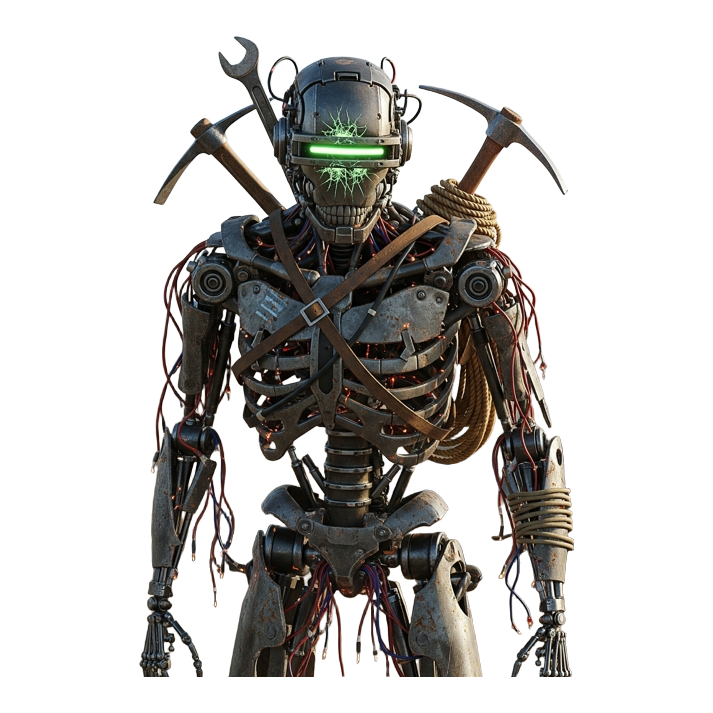} &
    \includegraphics[height=0.15\textwidth]{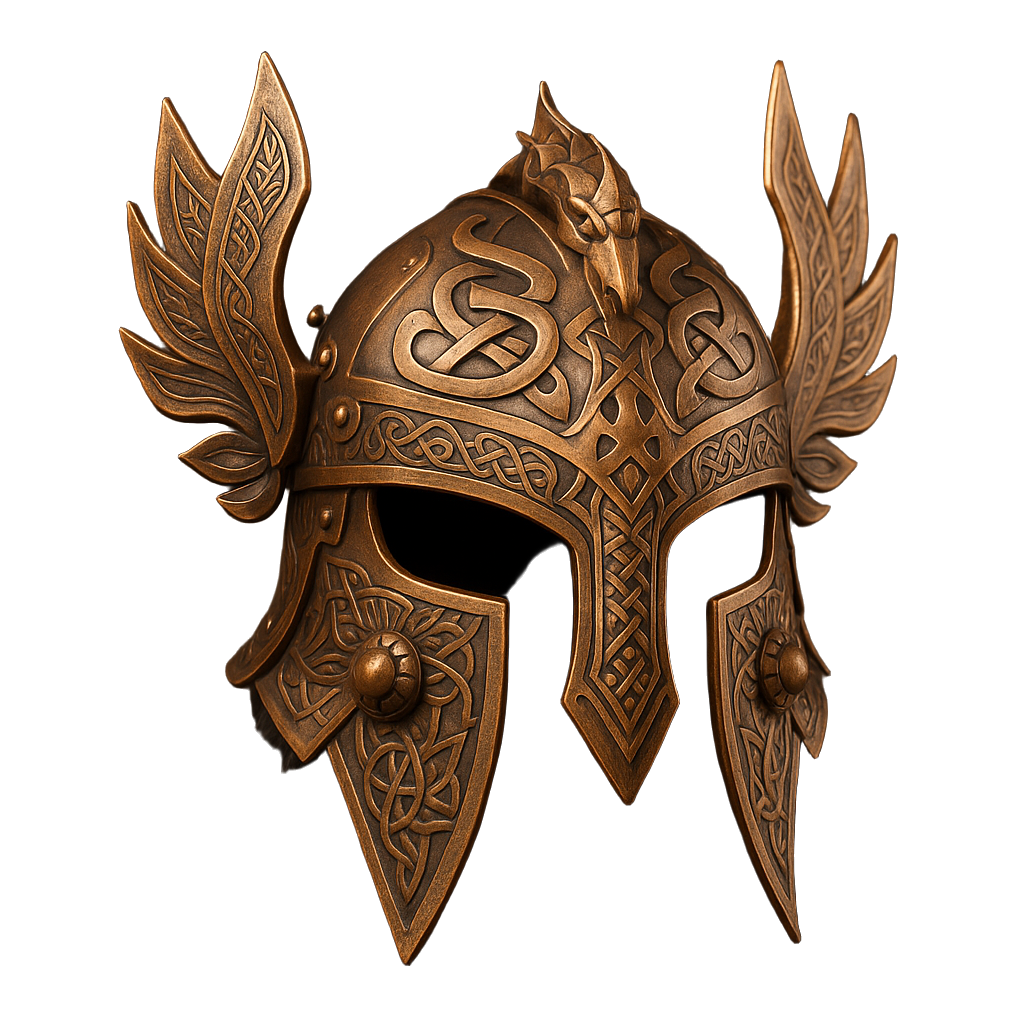} \\

    \rotatebox[origin=l]{90}{%
      \parbox[c]{0.15\textwidth}{%
        \centering
        Predicted Camera
      }%
    } &
    \includegraphics[height=0.15\textwidth]{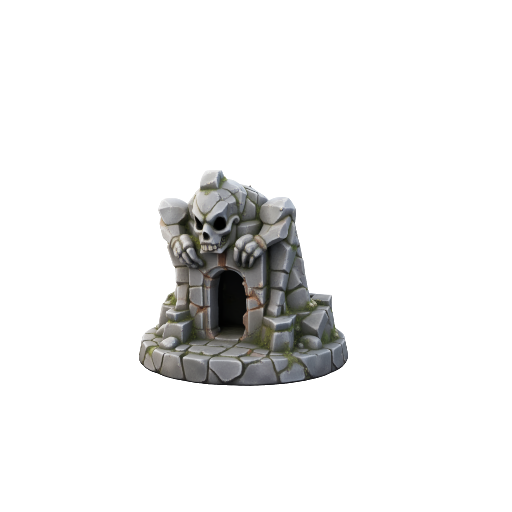} &
    \includegraphics[height=0.15\textwidth]{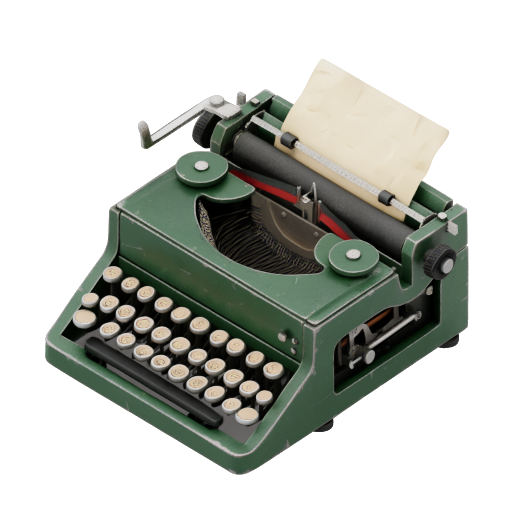} &
    \includegraphics[height=0.15\textwidth]{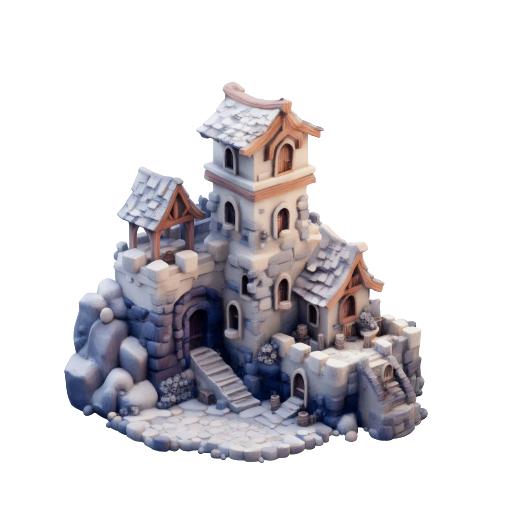} &
    \includegraphics[height=0.15\textwidth]{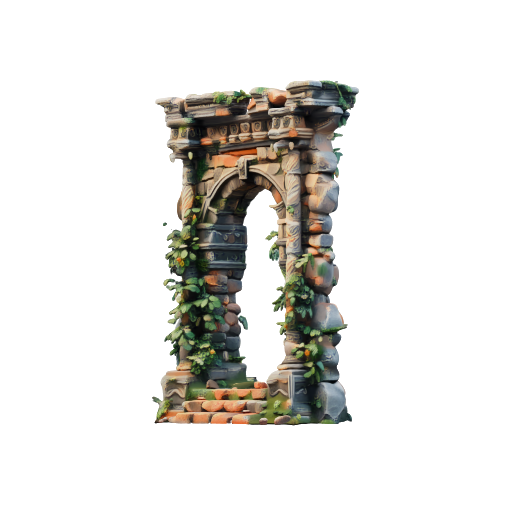} &
    \includegraphics[height=0.15\textwidth]{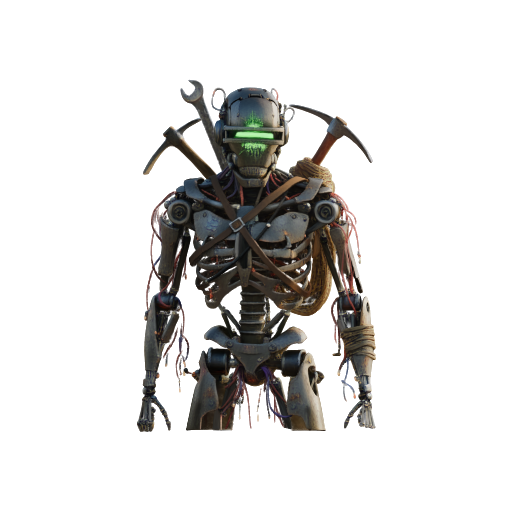} &
    \includegraphics[height=0.15\textwidth]{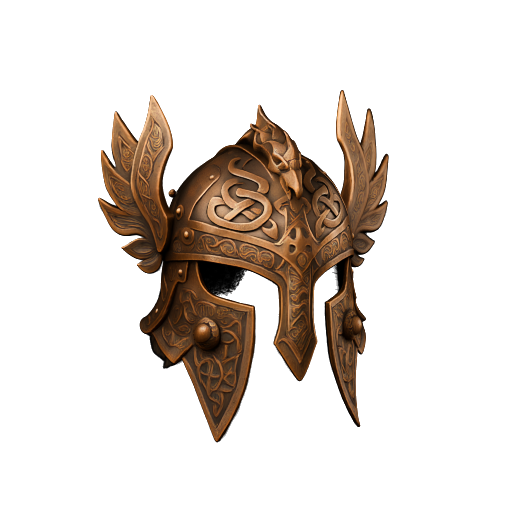} \\

  \end{tabular}%
  }
  \caption{Camera-token visualization for randomly selected examples from the test set. The first row shows the conditioning images. The second row shows the 3D Gaussians generated by the model, rendered using the predicted camera pose recovered from the camera token. The conditioning image is always scaled according to the alpha bounding box during both training and inference, so the predicted camera scale/distance may be inaccurate.}
  \label{fig:camera-vis}
\end{figure*}
\section{Failure Cases}

\newcounter{failurecase}
\renewcommand{\thefailurecase}{\alph{failurecase}}
\newsavebox{\failurecaseboxa}
\newsavebox{\failurecaseboxb}
Our model usually reconstructs the conditioning image well, even though we do not explicitly impose a dedicated loss or inductive bias for condition-image alignment. However, the synthesized unseen views can still fail in some cases, likely due to limited generative capacity. Fig.~\ref{fig:failure-cases} shows two representative examples. In Fig.~\ref{fig:failure-cases}(\ref{fig:failure-case-a}), the generated person matches the conditioning image well in the reference view, but the back view collapses to an unrealistic dark appearance. In Fig.~\ref{fig:failure-cases}(\ref{fig:failure-case-b}), the generated result is again highly consistent with the conditioning image in the reference view, while the frontal view contains an implausible head structure. We believe these failures may arise from two factors. First, the conditioning image itself is typically produced by a generative model and may not always depict a physically plausible 3D structure. Second, our model may still lack sufficient capacity to infer a fully plausible and realistic 3D structure from the input.

\begin{figure}[t]
  \centering
  \setlength{\tabcolsep}{1pt}
  \renewcommand{\arraystretch}{0.95}
  \begin{tabular}{cccc}
    \parbox[b][0.275\linewidth][c]{0.1\linewidth}{\small\refstepcounter{failurecase}(\thefailurecase)\label{fig:failure-case-a}} &
    \includegraphics[width=0.275\linewidth]{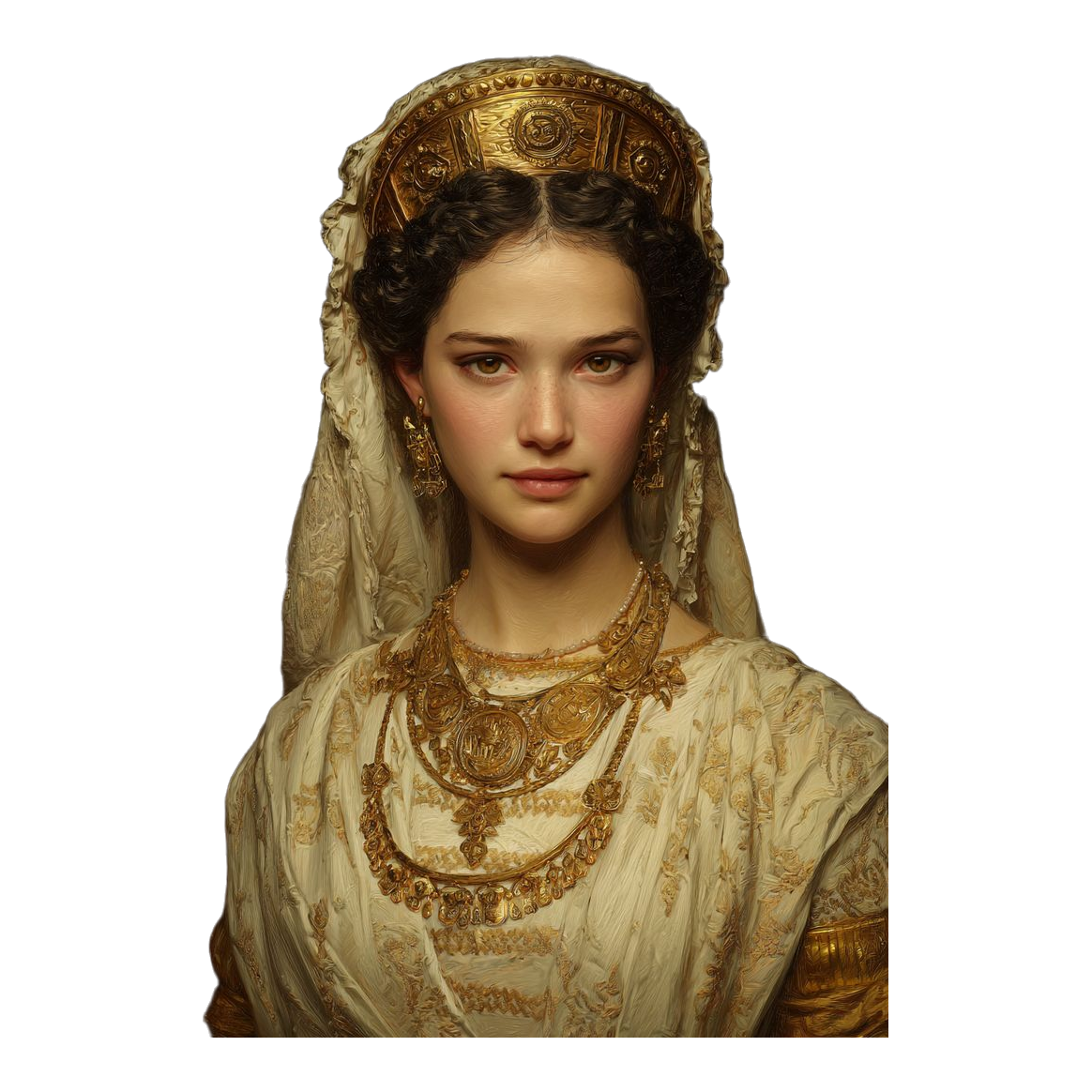} &
    \includegraphics[width=0.275\linewidth]{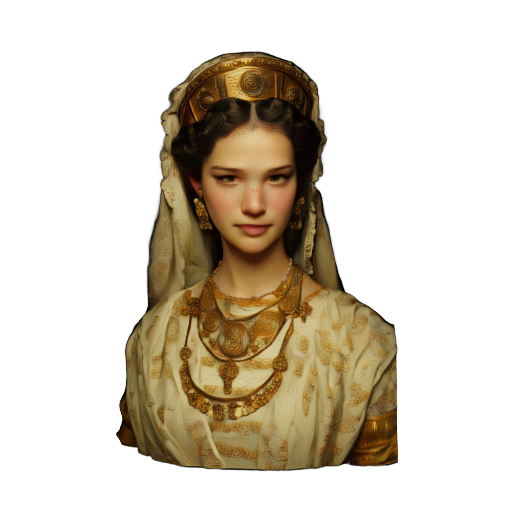} &
    \includegraphics[width=0.275\linewidth]{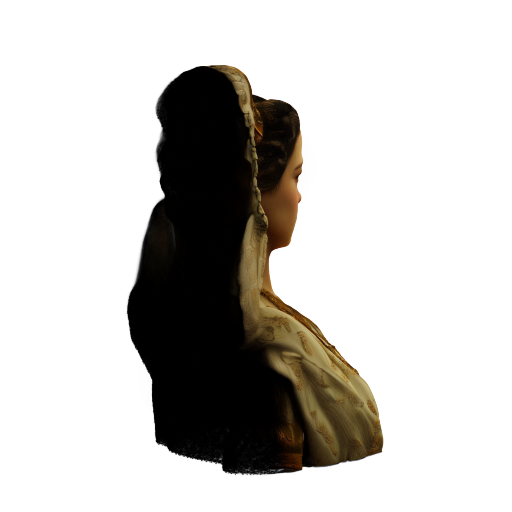} \\
    \parbox[b][0.275\linewidth][c]{0.1\linewidth}{\small\refstepcounter{failurecase}(\thefailurecase)\label{fig:failure-case-b}} &
    \includegraphics[width=0.275\linewidth]{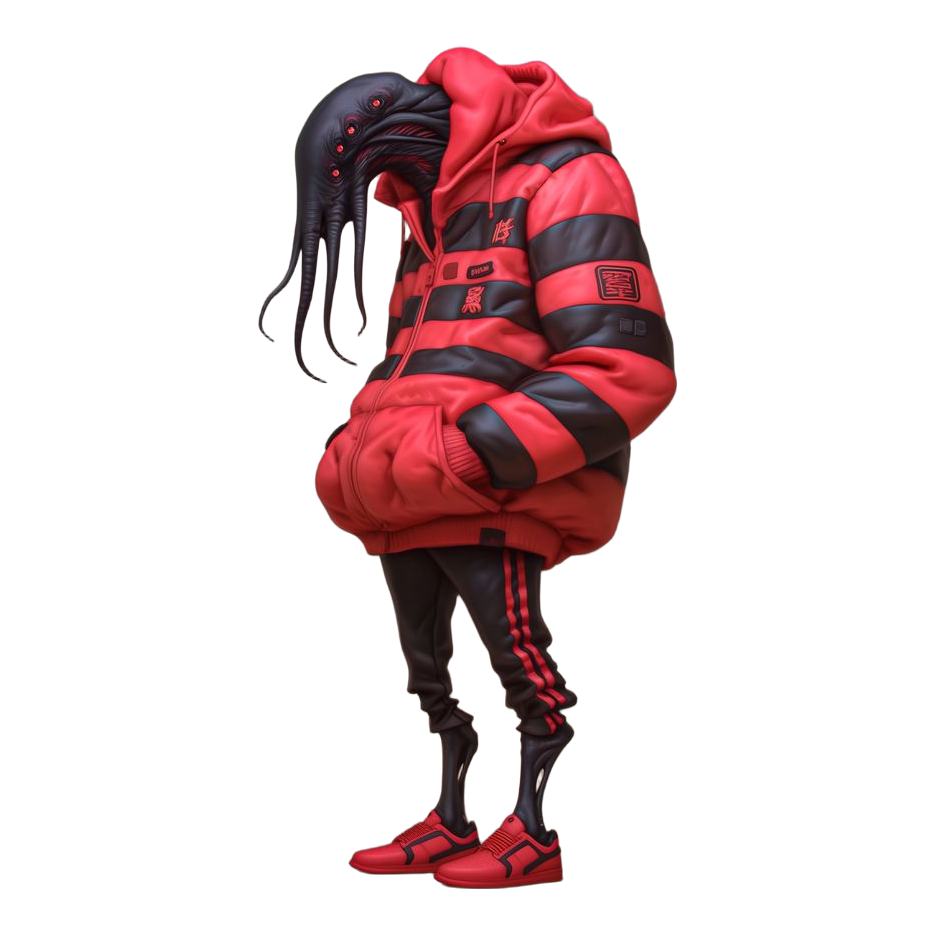} &
    \includegraphics[width=0.275\linewidth]{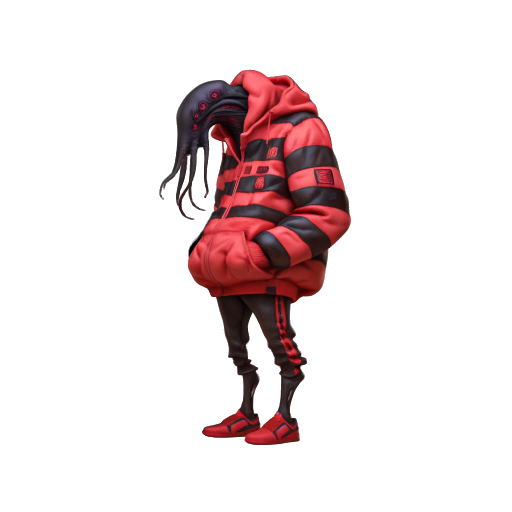} &
    \includegraphics[width=0.275\linewidth]{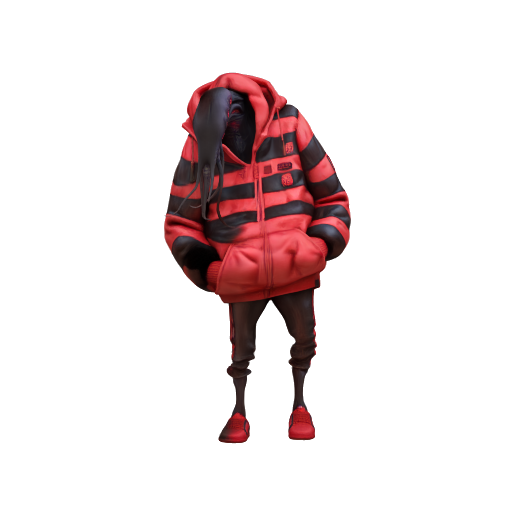} \\
    & \small Condition & \small Pred. Reference View & \small Pred. Side View \\
  \end{tabular}
  \caption{Representative failure cases of our conditional 3D Gaussian generation model. Each row shows one example. The model usually preserves strong consistency with the conditioning image in the reference view, but it may fail to synthesize plausible unseen-view geometry and appearance, leading to artifacts such as dark back views or structurally implausible results.}
  \label{fig:failure-cases}
\end{figure}

\section{Additional Comparison}
We provide additional visualizations of the generation results from our method, comparing them against the TRELLIS and TRELLIS.2 baselines (the leading generative baselines for GS and mesh representations, respectively). As shown in Fig.~\ref{fig:gen-comparison} and Fig.~\ref{fig:gen-comparison2}, our approach outperforms these baselines by exhibiting better condition alignment, richer details, and more natural colors.

\paragraph{User Study.}
To further validate the perceptual quality of our method, we conducted a user study focusing on complex prompts. We collected 94 challenging image prompts and generated the corresponding 3D assets using our method (DeG), TRELLIS, TRELLIS.2, UniLat3D, and Hunyuan3D 2.1. The generated assets were rendered as videos to provide a comprehensive 3D view. 

In the study, 32 anonymous participants were presented with 399 pairwise comparisons. In each comparison, participants viewed the same conditioning image alongside rendered videos from two different methods and were asked to choose the one with better overall quality and condition alignment. We computed Elo ratings based on these pairwise preferences. The results, summarized in Table~\ref{tab:user-study}, demonstrate that our method achieved a significant preference margin.

\begin{table*}[h]
  \centering
  \caption{User study results on 94 complex prompts. Elo ratings are computed from 399 pairwise comparisons by 32 participants. Higher Elo indicates stronger user preference.}
  \label{tab:user-study}
  \begin{tabular}{lccccc}
    \toprule
    Method & DeG (Ours) & TRELLIS & TRELLIS.2 & UniLat3D & Hunyuan3D 2.1 \\
    \midrule
    Elo $\uparrow$ & \textbf{1137} & 975 & 992 & 900 & \underline{996} \\
    \bottomrule
  \end{tabular}
\end{table*}

\section{Additional Ablations}

\paragraph{Additional Visualizations.}
We visualize the generated Gaussians with different Gaussian budgets from $33$K to $262$K in Fig.~\ref{fig:gaussian-count-vis}. Visual quality improves as the Gaussian budget increases.

\begin{figure*}[t]
  \centering
  \includegraphics[width=0.95\linewidth]{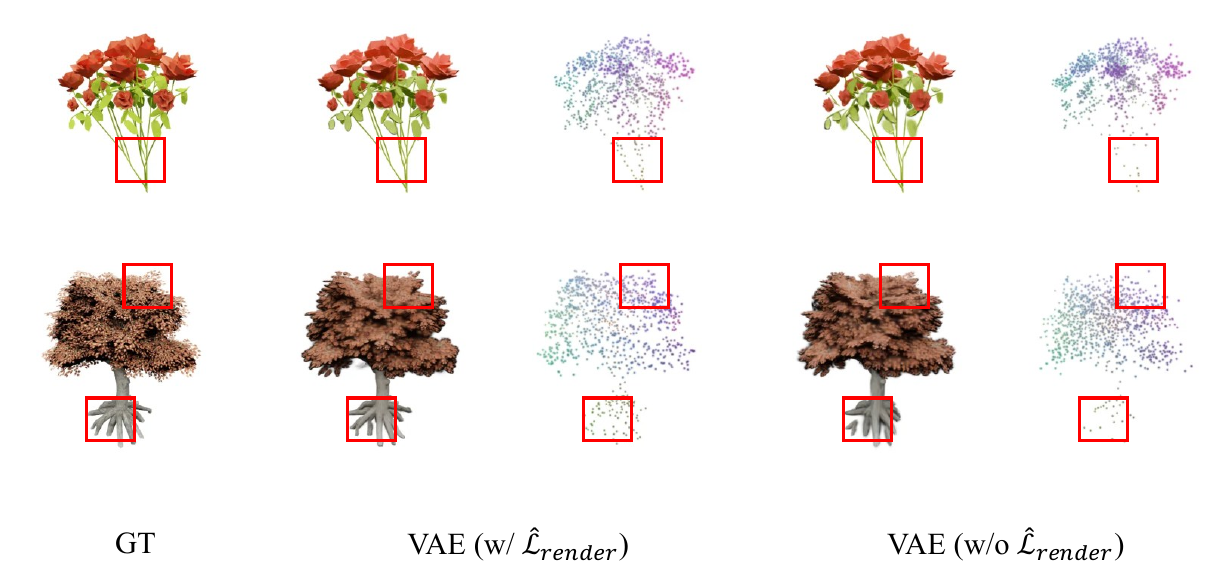}
  \caption{Visual comparison of VAE reconstruction with and without learned density control ($\Lrenderbp$). The Gaussian budget is set to $33$K. The anchor point clouds are shown alongside the rendered images. Red squares highlight regions where learned density control successfully allocates more Gaussian anchors to preserve fine details and prevent missing structures.}
  \label{fig:ablation-density}
\end{figure*}

\paragraph{Learned Density Control Visualizations.}
We compare the VAE reconstruction results with and without learned density control ($\Lrenderbp$) in Fig.~\ref{fig:ablation-density}. To validate the effectiveness of optimizing density via rendering supervision with $\Lrenderbp$, we train a Stage-3 VAE variant that disables $\Lrenderbp$ while keeping all other settings and training steps fixed. Incorporating $\Lrenderbp$ allocates more Gaussian anchors to complex regions, improving fine details and avoiding missing parts. As highlighted by the red squares, the anchor point clouds show that the model without learned density control fails to allocate sufficient capacity to thin structures and intricate details, leading to blurred or fragmented geometry, whereas our full model preserves these features cleanly. This visualization complements the quantitative discussion of learned density control provided in the main text.

\begin{figure}[t]
  \centering
  \begin{tabular}{cc}
    \includegraphics[width=0.48\linewidth]{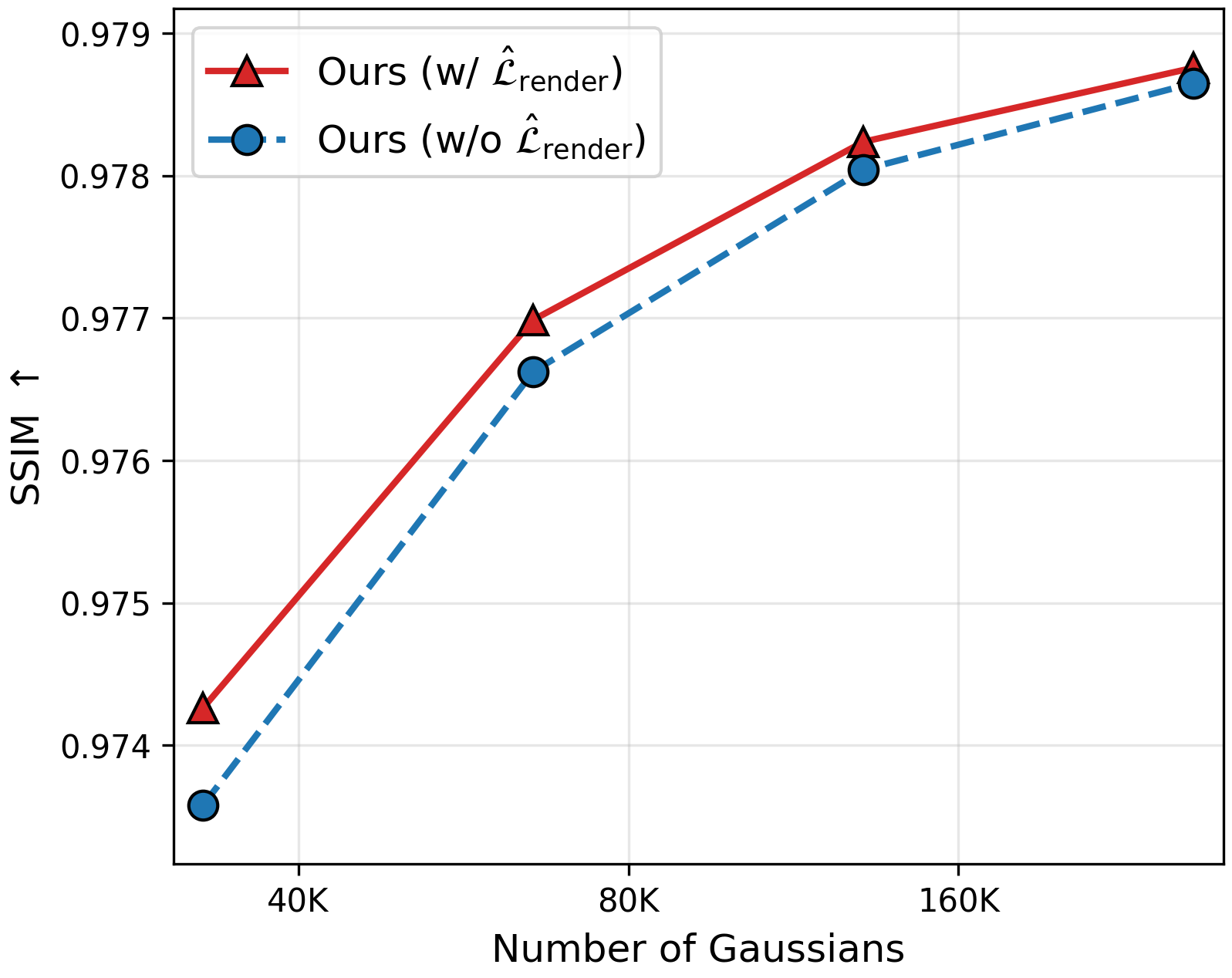} &
    \includegraphics[width=0.48\linewidth]{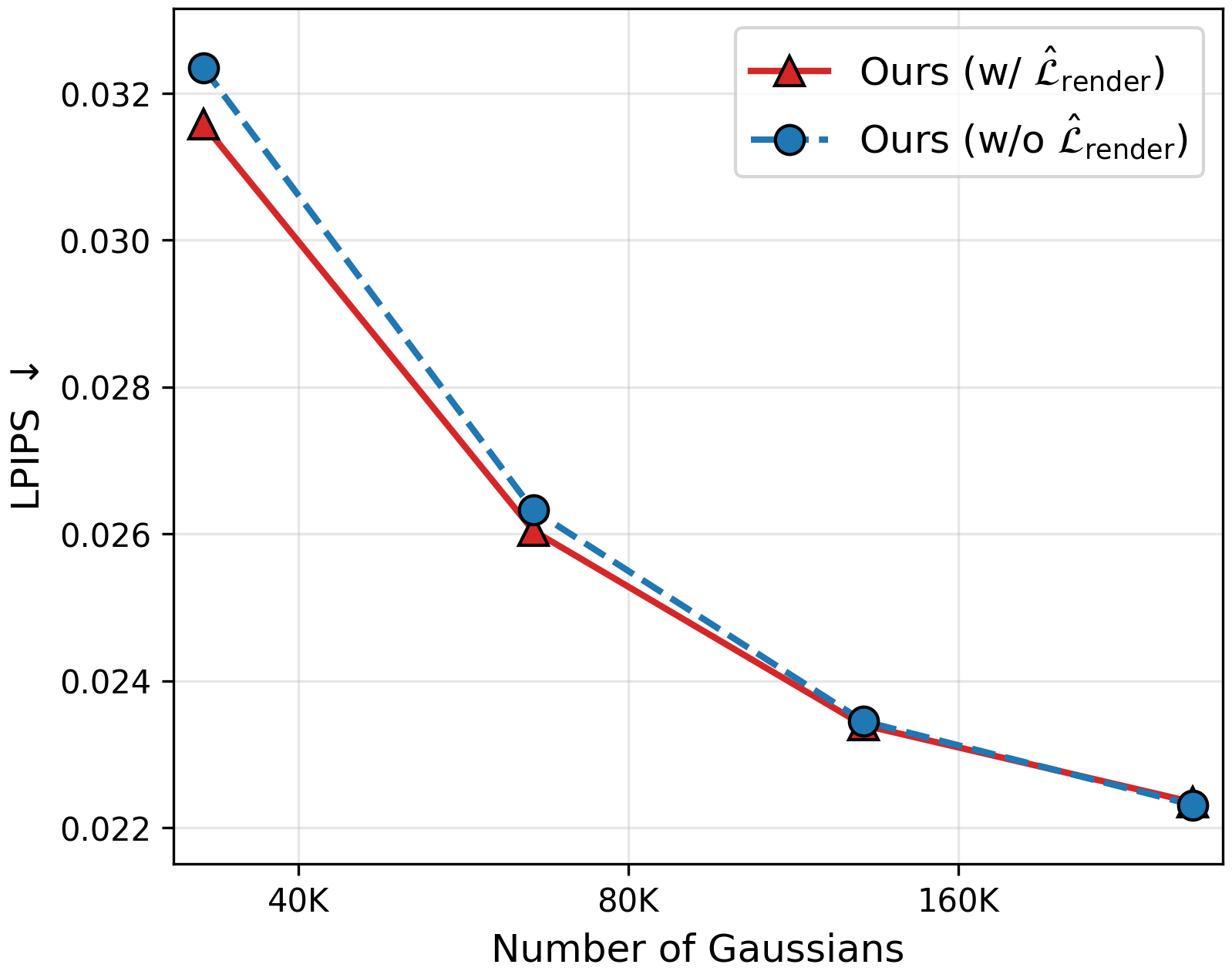} \\
    \small (a) & \small (b)
  \end{tabular}
  \caption{Additional metrics for the learned density control setting in the main text under different Gaussian budgets: (a) SSIM$\uparrow$ and (b) LPIPS$\downarrow$.}
  \label{fig:ablation-ssim-lpips}
\end{figure}

\paragraph{Hyperparameter Ablations.}
We ablate key hyperparameters of \ourgsvae, fine-tuned from the Stage~1 checkpoint for 100K steps ($P=2048$, batch size~2). Results are reported in Table~\ref{tab:vae-ablation}.
For the local expansion factor $K$, $K=32$ offers a good trade-off; increasing to $K=64$ brings only marginal PSNR gain at higher cost.
For the octree depth $L$, $L=8$ achieves the best PSNR; $L=10$ yields comparable quality but incurs higher training time.
Removing $\Lreg$ slightly reduces PSNR. 
%
\paragraph{Additional Metrics for Learned Density Control.}
We additionally report SSIM and LPIPS for the same learned density control setting as in the main text; Fig.~\ref{fig:ablation-ssim-lpips} visualizes the corresponding trends. Compared with the variant without $\Lrenderbp$, enabling $\Lrenderbp$ improves both SSIM and LPIPS in the low-budget regime, while the gap becomes small at larger Gaussian budgets, unlike PSNR. A possible explanation is that the render-loss contribution gradient is derived only from the $\mathcal{L}_{\text{l1}}$ term and may therefore underfit perceptual errors in the high-budget regime.

\begin{table}[t]
  \centering
  \caption{Hyperparameter ablations of \ourgsvae (PSNR$\uparrow$, SSIM$\uparrow$, LPIPS$\downarrow$ on held-out validation set from the training data, training time in GPU-hours). All variants fine-tuned from the Stage~1 checkpoint for 100K steps with $P=2048$. $^\dagger$ marks the setting used in the main experiments and final training.}
  \label{tab:vae-ablation}
  \setlength{\tabcolsep}{6pt}
  \begin{tabular}{lcccc}
    \toprule
    Variant & PSNR$\uparrow$ & SSIM$\uparrow$ & LPIPS$\downarrow$ & Time (h) \\
    \midrule
    $K=8$  & 
    27.39 & 0.9383 & 0.0887 & 47.0 \\
    $K=16$ & 
    27.40 & 0.9389 & 0.0863 & 47.7 \\
    $K=32^\dagger$ & 
    \underline{27.48} & \underline{0.9398} & \underline{0.0835} & 47.0 \\
    $K=64$ & 
    \textbf{27.49} & \textbf{0.9401} & \textbf{0.0812} & 47.8 \\
    \midrule
    $L=4$  & 
    27.17 & 0.9381 & \underline{0.0836} & 45.0 \\
    $L=6$  & 
    27.10 & 0.9378 & 0.0839 & 43.5 \\
    $L=8^\dagger$ & 
    \textbf{27.48} & \textbf{0.9398} & \textbf{0.0835} & 47.0  \\
    $L=10$ & 
    \underline{27.46} & \underline{0.9396} & \underline{0.0836} & 52.6 \\
    \midrule
    w/o $\Lreg$ & 
    27.44 & 0.9397 & \textbf{0.0834} & 46.7 \\
    w/ $\Lreg^\dagger$ & 
    \textbf{27.48} & \textbf{0.9398} & 0.0835 & 47.0 \\
    \bottomrule
  \end{tabular}
\end{table}

\section{Regularization}
\label{sec:appendix:details}
In this appendix, we provide additional details for the \ourgsvae training objective, specifically the regularization term $\Lreg$. $\Lreg$ comprises three terms:
\begin{equation}
  \Lreg = \lambda_1 \cdot \frac{1}{N} \sum_{i=1}^N \prod_{d=1}^3 \Sigma_i^d + \lambda_2 \cdot \frac{1}{N} \sum_{i=1}^N (1-\alpha_i) + \lambda_\text{offset} \Loffset.
\end{equation}
The first two terms regularize the volume and opacity of each 3D Gaussian, respectively, encouraging compact and sparse primitives.
The third term $\Loffset$ enforces the spatial compactness and separation of the Gaussian clusters decoded from each anchor. We treat the set of Gaussians generated from anchor $x_i$ as a cluster with offsets $\{\delta_{i,k}\}_{k=1}^K$. We define the cluster spread $\sigma_i = \sqrt{\frac{1}{K} \sum_{k=1}^K \|\delta_{i,k}\|^2}$ and mean offset $\bar{\delta}_i = \frac{1}{K} \sum_{k=1}^K \delta_{i,k}$. The loss consists of two components:
\begin{equation}
  \Loffset = \Loffset^\text{center} + \Loffset^\text{sep}.
\end{equation}
The centering term $\Loffset^\text{center}$ ensures the cluster remains centered around its anchor:
\begin{equation}
  \Loffset^\text{center} = \frac{1}{P} \sum_{i=1}^P \operatorname{ReLU}(\|\bar{\delta}_i\| - \gamma \sigma_i),
\end{equation}
where $\gamma=0.5$ is a hyperparameter controlling the permissible drift. The separation term $\Loffset^\text{sep}$ prevents cluster overlap by constraining the spread to be smaller than the distance to other anchors:
\begin{equation}
  \Loffset^\text{sep} = \frac{1}{P^2} \sum_{i=1}^P \sum_{j \neq i}^P \operatorname{ReLU}(\sigma_i - \|x_i - x_j\|).
\end{equation}

\section{Octree Sampling Algorithm}
\label{sec:appendix:sampling}
We provide our implementation of the efficient octree sampling algorithm in Alg.~\ref{alg:sampling}. We use this algorithm to sample $P$ anchor points from \probdecodertext $\probdecoder$ at inference time.

\section{Primitive-Level Implementation of the \lonec}

\paragraph{Setup.}
Following the main paper, we first sample $P$ anchors, and each anchor $x_a$ spawns $K$ Gaussian primitives $\{g_{a,k}\}_{k=1}^{K}$ via local expansion. We extend the rasterizer used in Mip-Splatting~\cite{yu2024mip} with \lonec computation. The rasterizer itself does not operate on anchors; it only receives the flattened primitive set
\begin{equation}
  \gs = \{g_i\}_{i=1}^{N}, \quad N = P \cdot K,
\end{equation}
where each flattened index $i$ corresponds to some anchor-primitive pair $(a,k)$. Let
\begin{equation}
  I = \mathcal{R}(\gs, \pi) \in \mathbb{R}^{C \times H \times W}
\end{equation}
be the rendered image. In this section, we focus on the $\mathcal{L}_\text{l1}$ term inside the render loss $\Lrender$. During training, all $K$ primitives generated from the same anchor share the same anchor probability $\probdecoder(x_a \mid \latent)$ when constructing the density-learning signal. The residual at pixel $p$ in channel $c$ is
\begin{equation}
  R_{p,c} = I_{p,c} - I^{\mathrm{GT}}_{p,c},
\end{equation}
computed once after the forward pass and stored for the backward pass.

\paragraph{Target Quantity.}
For each Gaussian primitive $g_i$ we want the quantity
\begin{equation}
  \Delta \mathcal{L}_{\text{l1},i} = \mathcal{L}_\text{l1}(I,\, I^{\mathrm{GT}}) - \mathcal{L}_\text{l1}(I^{(-i)},\, I^{\mathrm{GT}}),
  \label{eq:delta_L}
\end{equation}
where $I^{(-i)}$ is the image that would be rendered if $g_i$ were absent.
Computing~\eqref{eq:delta_L} naively requires $N$ additional forward passes. Instead, we derive an efficient implementation that computes this quantity for all Gaussians \textbf{within a standard forward and backward pass}, with negligible overhead. The resulting contributions are then associated with the shared anchor probability of the source anchor.

\paragraph{Color change from removing one primitive.}
Standard alpha compositing renders pixel $p$ as
\begin{equation}
  I_{p,c} = \sum_i T_{p,i}\,\alpha_{p,i}\,c_{i,c} + T_{p,\text{final}}\,b_c,
\end{equation}
where $T_{p,i} = \prod_{j < i}(1 - \alpha_{p,j})$ is the transmittance just before primitive $i$, $\alpha_{p,i}$ is its blending weight, $c_{i,c}$ is its color, and $b_c$ is the background color.

If $g_i$ is removed, the pixel color changes by
\begin{equation}
  \Delta C_{p,i,c} = T_{p,i}\,\alpha_{p,i}\,(c_{i,c} - \mathrm{back}_{p,i,c}),
  \label{eq:delta_c}
\end{equation}
In a standard backward pass, both the transmittance $T_{p,i}$ and the back color $\mathrm{back}_{p,i,c}$ are already accumulated, so this color change can be computed efficiently for every primitive at negligible extra cost. Here, $\mathrm{back}_{p,i,c}$ denotes the blended color already accumulated behind $g_i$ in the original rasterizer, i.e.\ the color that would be seen from layers deeper than $g_i$.

\paragraph{Per-pixel L1 change.}
The change in the per-pixel, per-channel L1 loss when $g_i$ is removed is
\begin{equation}
  \delta^{\text{l1}}_{p,i,c} = |R_{p,c}| - |R_{p,c} - \Delta C_{p,i,c}|.
  \label{eq:delta_l1}
\end{equation}
Summing over all pixels and channels gives the primitive-level contribution of $g_i$:
\begin{equation}
  \Delta \mathcal{L}_{\text{l1},i} = \sum_p \sum_c \delta^{\text{l1}}_{p,i,c}.
  \label{eq:accum}
\end{equation}

\paragraph{Fused CUDA implementation.}
The standard backward pass already iterates over all (primitive, pixel) pairs in reverse depth order to compute gradients with respect to Gaussian parameters.
We accumulate $\Delta \mathcal{L}_{\text{l1},i}$ inside this same loop at negligible extra cost.
At each step the backward pass maintains the transmittance $T_{p,i}$ (unwound from the final transmittance) and the accumulated background color $\mathrm{back}_{p,i,c}$ (built up from the back of the scene).
These two quantities are exactly what Eq.~\eqref{eq:delta_c} requires, so computing $\Delta C_{p,i,c}$ and the resulting $\delta^{\text{l1}}_{p,i,c}$ adds only a few arithmetic operations per (primitive, pixel, channel) triple.
The per-primitive sum $\Delta \mathcal{L}_{\text{l1},i}$ is accumulated via an atomic add into a buffer of length $N$. Outside the rasterizer, each primitive contribution is paired with the probability of its source anchor; hence all $K$ primitives expanded from the same anchor reuse the same factor $\probdecoder(x_a \mid \latent)$ in the density-learning objective.

\paragraph{Reward Clamping.}
In our experiments, we apply a simple clamping rule to improve the performance of the density-learning signal from the \lonec.
After accumulating anchor-level contributions, we apply two clamping operations.
First, we compute the $10$th-percentile threshold and clamp the smallest $10\%$ of contributions to this threshold value.
Second, we clamp positive contributions, i.e., contributions with $\lonec>0$, to zero.
This keeps the update focused on locations where the presence of an anchor leads to a larger reduction in rendering loss, while avoiding high-variance outliers.

\paragraph{Scope and Limitation.}
This implementation only considers the $\mathcal{L}_\text{l1}$ component of the render loss $\Lrender$. Perceptual terms such as LPIPS are not pixel-wise losses, so assigning an exact per-primitive contribution inside the rasterizer is difficult. Therefore, this section only computes the contribution with respect to the main rendering $\mathcal{L}_\text{l1}$ term.

\section{Effect of Token Count and Gaussian Budget on VAE Reconstructions}

We provide a qualitative comparison of VAE reconstructions for two controlled sweeps.
In the first sweep, we vary the token length while fixing the Gaussian budget to $262$K.
In the second sweep, we vary the Gaussian budget while fixing the token length to $8192$. We report PSNR$\uparrow$, SSIM$\uparrow$, and LPIPS$\downarrow$ on the Toys4K dataset in Table~\ref{tab:vae-recon-token-gaussian}.

\begin{table}[t]
  \centering
  \caption{Quantitative VAE reconstruction results for the tested token-length and Gaussian-budget sweeps. We vary token length with a fixed Gaussian budget of $262$K, and vary Gaussian budget with a fixed token length of $8192$.}
  \label{tab:vae-recon-token-gaussian}
  \setlength{\tabcolsep}{3pt}
  \renewcommand{\arraystretch}{1.05}
  \textbf{(a)} Token-length sweep with fixed Gaussian budget $262$K.
  \begin{tabular}{c|ccc}
    \toprule
    Token length & PSNR$\uparrow$ & SSIM$\uparrow$ & LPIPS$\downarrow$ \\
    \midrule
    $1024$ &34.94 & 0.9750 & 0.0265 \\
    $2048$ &35.46 & 0.9770 & 0.0243 \\
    $4096$ &35.74 & 0.9781 & 0.0230 \\
    $8192$ &\textbf{35.89} & \textbf{0.9788} & \textbf{0.0223} \\
    \bottomrule
  \end{tabular}
\vspace{4pt}

  \textbf{(b)} Gaussian-budget sweep with fixed token length $8192$.
  \begin{tabular}{c|ccc}
    \toprule
    \# Gaussians & PSNR$\uparrow$ & SSIM$\uparrow$ & LPIPS$\downarrow$ \\
    \midrule
    $33$K & 34.73 & 0.9743 & 0.0316 \\
    $66$K & 35.45 & 0.9770 & 0.0260 \\
    $131$K &35.77 & 0.9782 & 0.0234 \\
    $262$K &\textbf{35.89} & \textbf{0.9788} & \textbf{0.0223} \\
    \bottomrule
  \end{tabular}
\end{table}


\newlength{\budgetimgw}
\setlength{\budgetimgw}{0.15\textwidth}
\newlength{\cropimgw}
\setlength{\cropimgw}{0.065\textwidth}

\newcommand{\budgetcropsA}[1]{%
  \includegraphics[width=\cropimgw,trim={140pt 220pt 290pt 210pt},clip]{#1}\hspace{1pt}%
  \includegraphics[width=\cropimgw,trim={250pt 360pt 170pt 80pt},clip]{#1}\hspace{1pt}%
  \includegraphics[width=\cropimgw,trim={250pt 180pt 170pt 250pt},clip]{#1}%
}
\newcommand{\budgetcropsB}[1]{%
  \includegraphics[width=\cropimgw,trim={140pt 240pt 290pt 190pt},clip]{#1}\hspace{1pt}%
  \includegraphics[width=\cropimgw,trim={250pt 140pt 170pt 290pt},clip]{#1}\hspace{1pt}%
  \includegraphics[width=\cropimgw,trim={250pt 270pt 170pt 160pt},clip]{#1}%
}

\newcommand{\budgetrowcat}{%
  \includegraphics[width=\budgetimgw]{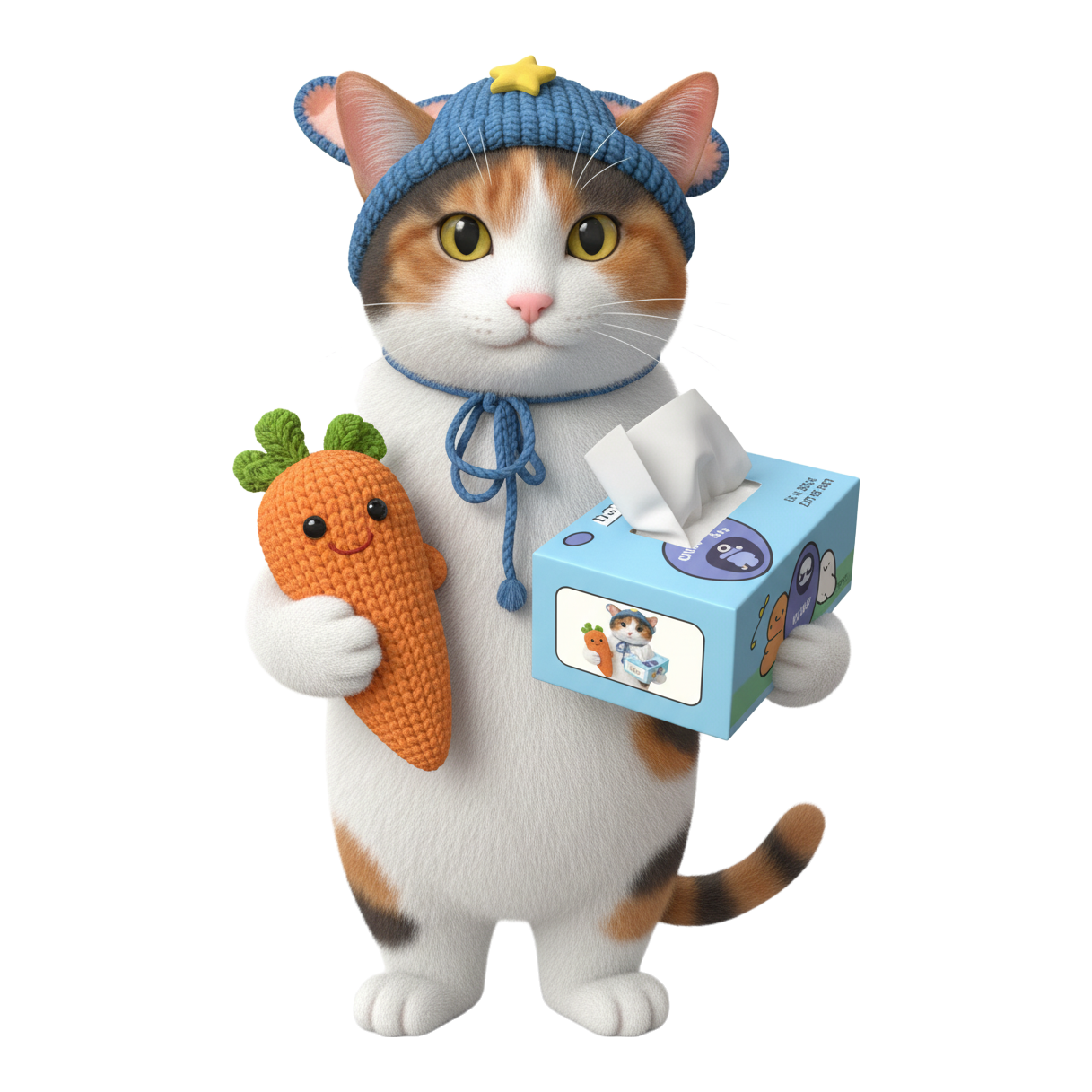} &
  \includegraphics[width=\budgetimgw]{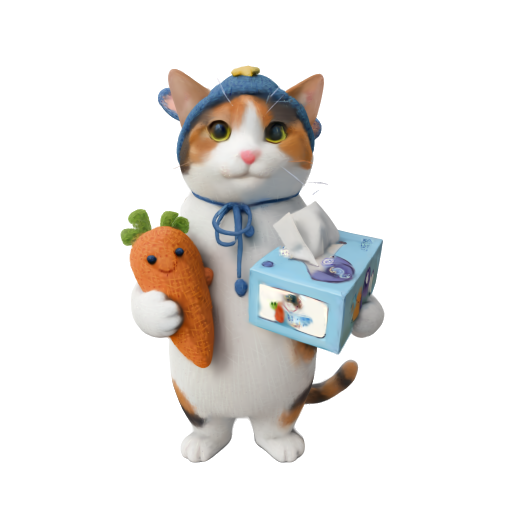} &
  \includegraphics[width=\budgetimgw]{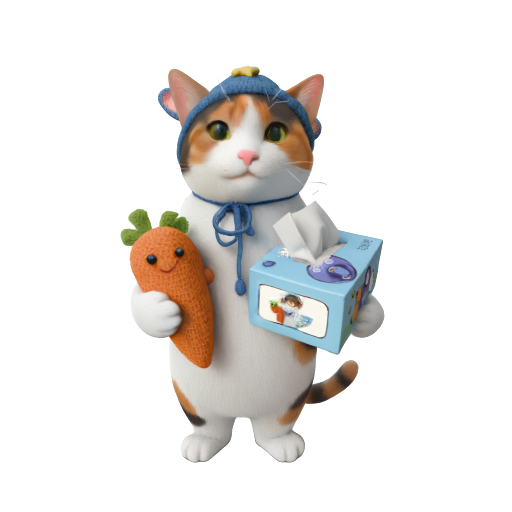} &
  \includegraphics[width=\budgetimgw]{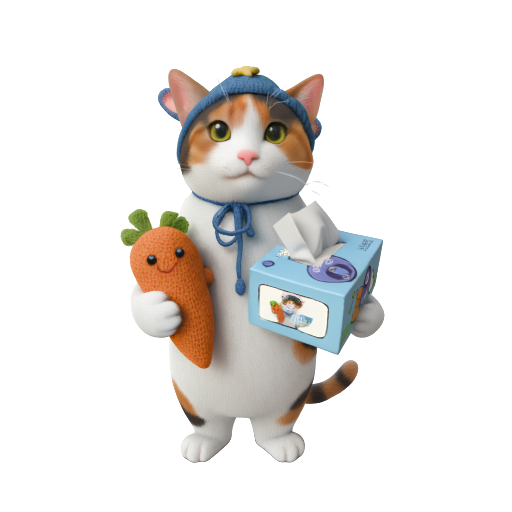} &
  \includegraphics[width=\budgetimgw]{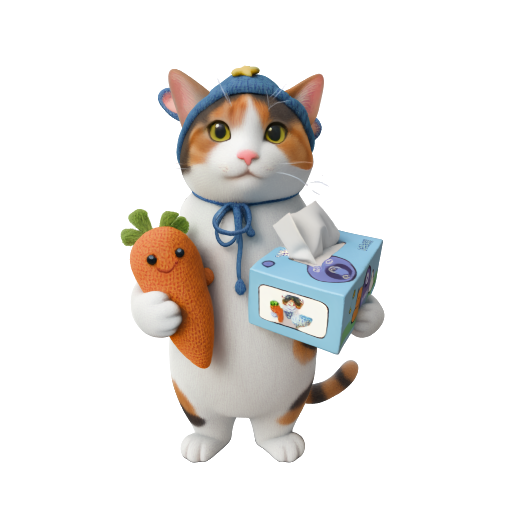} \\[1pt]
  &
  \budgetcropsA{figures/budget/cat/budget_01024.png} &
  \budgetcropsA{figures/budget/cat/budget_02048.png} &
  \budgetcropsA{figures/budget/cat/budget_04096.png} &
  \budgetcropsA{figures/budget/cat/budget_08192.png} \\[6pt]
}

\newcommand{\budgetrowtable}{%
  \includegraphics[width=\budgetimgw]{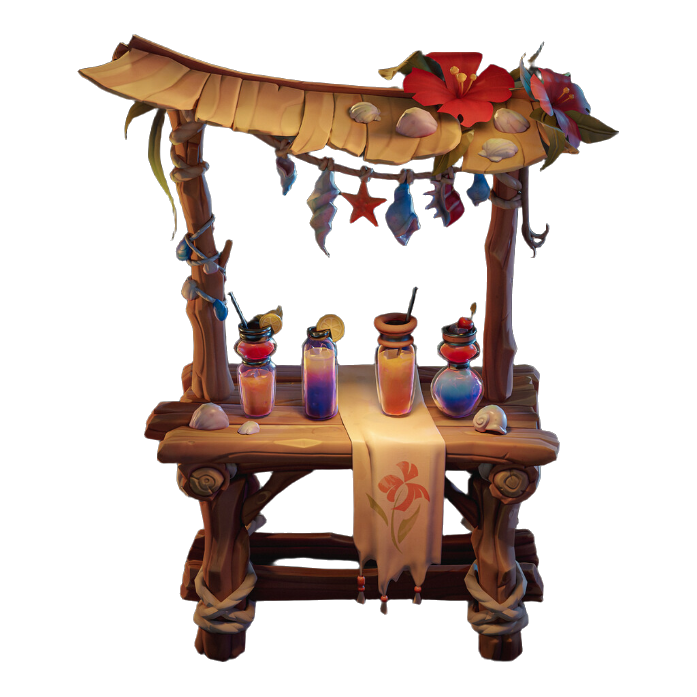} &
  \includegraphics[width=\budgetimgw]{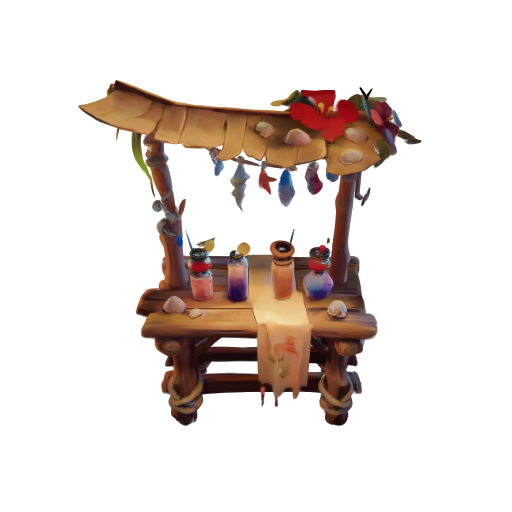} &
  \includegraphics[width=\budgetimgw]{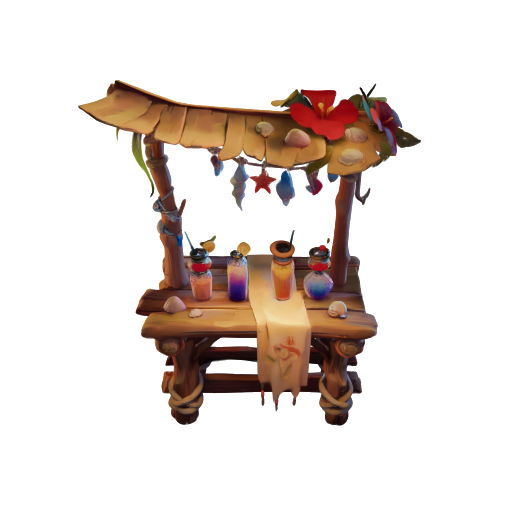} &
  \includegraphics[width=\budgetimgw]{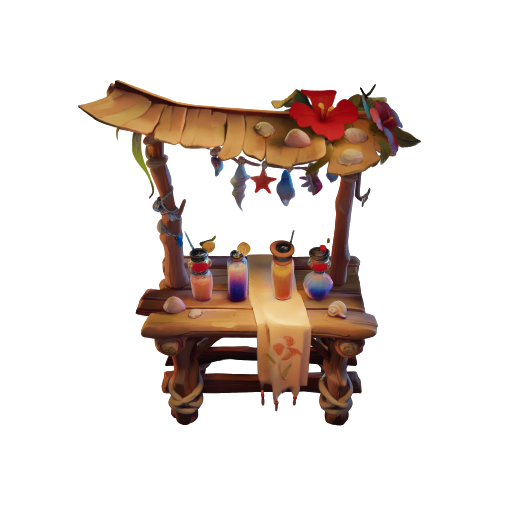} &
  \includegraphics[width=\budgetimgw]{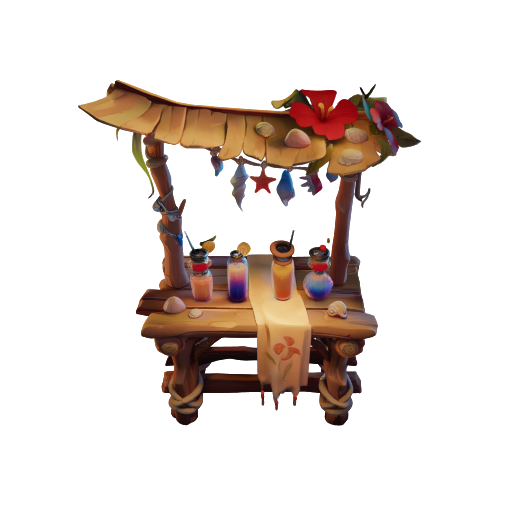} \\[1pt]
  &
  \budgetcropsB{figures/budget/table/budget_01024.png} &
  \budgetcropsB{figures/budget/table/budget_02048.png} &
  \budgetcropsB{figures/budget/table/budget_04096.png} &
  \budgetcropsB{figures/budget/table/budget_08192.png} \\[6pt]
}
\begin{figure*}[t]
  \centering
  \setlength{\tabcolsep}{1.5pt}
  \renewcommand{\arraystretch}{0.5}
  \begin{tabular}{ccccc}
    Condition & Budget=33K & Budget=66K & Budget=131K & Budget=262K \\[2pt]
    \budgetrowcat
    \budgetrowtable
  \end{tabular}
  \caption{Visualizations under different Gaussian budgets, ranging from $33$K to $262$K. For each rendering, three zoomed-in crops are shown beneath the full image. Each row corresponds to the same generated latent decoded with a different Gaussian budget.}
  \label{fig:gaussian-count-vis}
\end{figure*}

\newcommand{\genmethodcell}[1]{%
  \begin{minipage}{0.22\textwidth}
    \includegraphics[width=\linewidth]{#1/image/001.png}\\[2pt]
    \includegraphics[width=0.49\linewidth]{#1/image/000.png}\hfill
    \includegraphics[width=0.49\linewidth]{#1/image/002.png}
  \end{minipage}%
}
\newcommand{\gencondcell}[1]{%
  \begin{minipage}{0.22\textwidth}
    \includegraphics[width=\linewidth]{figures/gen-compair/#1.png}
  \end{minipage}%
}
\newcommand{\genfull}[1]{
  \gencondcell{#1} & \genmethodcell{figures/gen-compair/vecseq_#1} & \genmethodcell{figures/gen-compair/trellis_#1} & \genmethodcell{figures/gen-compair/trellis2_#1}
}
\begin{figure*}[h]
  \centering
  \begin{tabular}{cccc}
    Condition & DeG (Ours) & TRELLIS & TRELLIS.2 \\[4pt]
     \\[4pt]
    \genfull{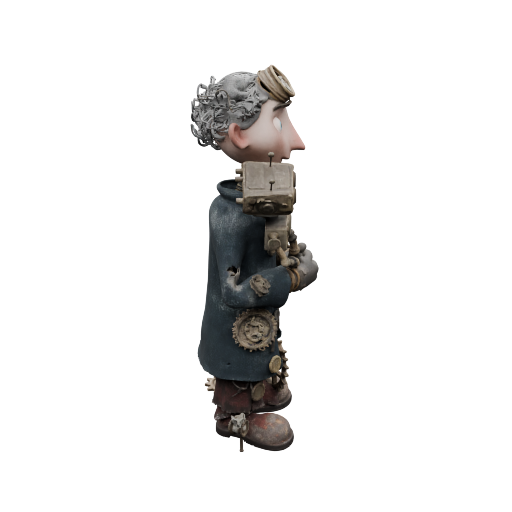}\\[4pt]
    \genfull{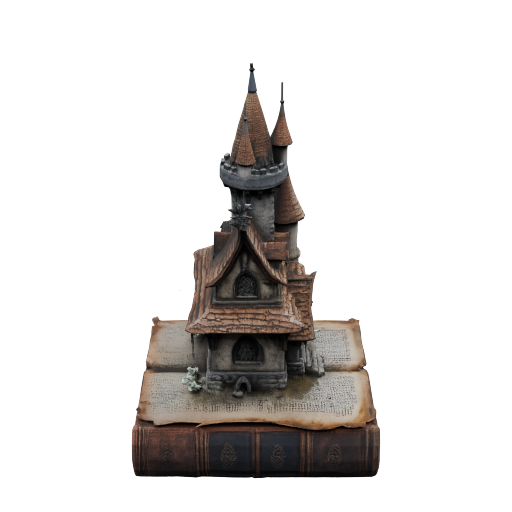}\\[4pt]
    \genfull{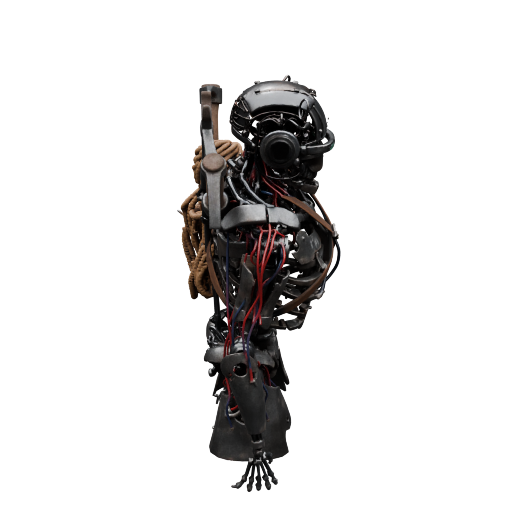}\\
  \end{tabular}
  \vspace{23pt} 
  \caption{Additional visualizations of the generation results. DeG and TRELLIS are shown with Gaussian rendering, while TRELLIS-2 is shown with mesh rendering. Our method demonstrates better condition alignment, richer details, and more natural colors compared with the baselines. The Gaussian count for our method (DeG) is $262$K. The Gaussian counts for TRELLIS are $360$K, $456$K, and $459$K from top to bottom, respectively.}
  \label{fig:gen-comparison}
\end{figure*}
\begin{figure*}[h]
  \centering
  \begin{tabular}{cccc}
    Condition & DeG (Ours) & TRELLIS & TRELLIS.2 \\[4pt]
     \\[4pt]
    \genfull{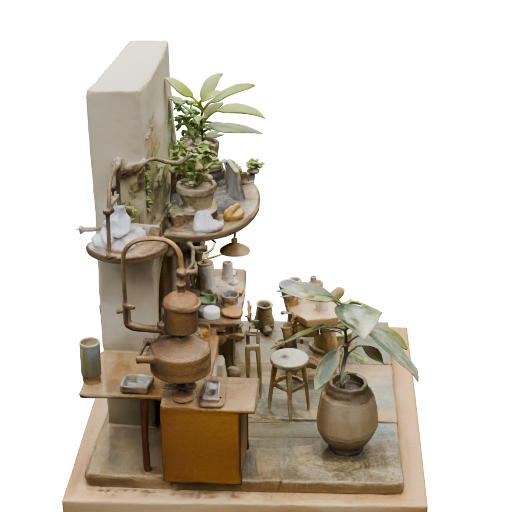}\\[4pt]
    \genfull{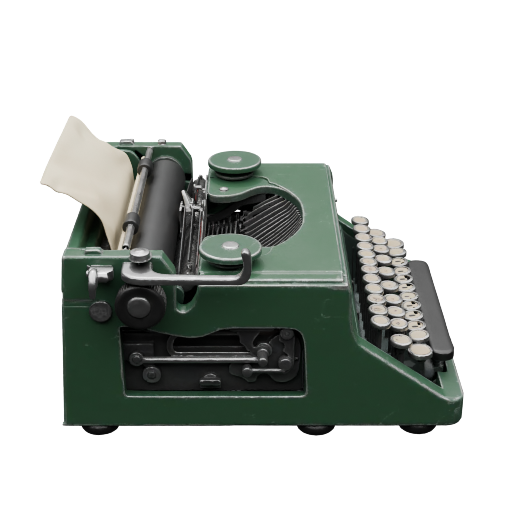}\\[4pt]
    \genfull{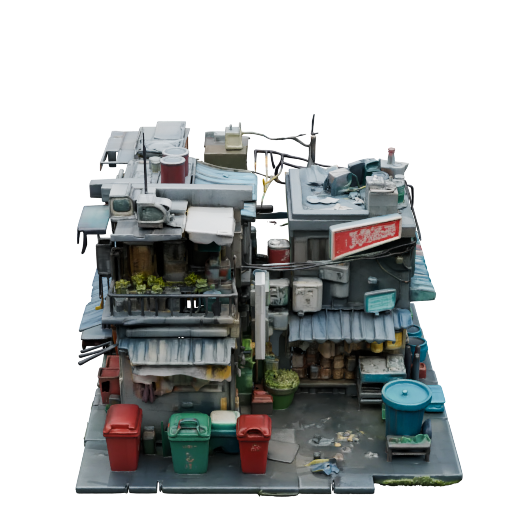}\\
  \end{tabular}
  \vspace{23pt} 
  \caption{Additional visualizations of the generation results. DeG and TRELLIS are shown with Gaussian rendering, while TRELLIS-2 is shown with mesh rendering. Our method demonstrates better condition alignment, richer details, and more natural colors compared with the baselines. The Gaussian count for our method (DeG) is $262$K. The Gaussian counts for TRELLIS are $776$K, $912$K, and $907$K from top to bottom, respectively.}
  \label{fig:gen-comparison2}
\end{figure*}

\end{document}